\documentclass[draft]{agujournal2019}
\usepackage{url} 
\usepackage[inline]{trackchanges} 
\usepackage{soul}
\usepackage{amsmath}
\usepackage{amssymb}

\usepackage{algorithm}
\usepackage{algpseudocode}

\usepackage{tikz}
\usetikzlibrary{matrix,
                positioning,
                quotes,
                cd}
\journalname{Journal of Advances in Modeling Earth Systems (JAMES)}

\begin{document}

%
%


\title{A posteriori learning for quasi-geostrophic turbulence parametrization}

%
%




\authors{Hugo Frezat\affil{1,2,3}, Julien Le Sommer\affil{2}, Ronan Fablet\affil{3}, Guillaume Balarac\affil{1,4} and Redouane Lguensat\affil{5}}

\affiliation{1}{Univ. Grenoble Alpes, CNRS UMR LEGI, Grenoble, France}
\affiliation{2}{Univ. Grenoble Alpes, CNRS UMR IGE, Grenoble, France}
\affiliation{3}{IMT Atlantique, CNRS UMR Lab-STICC, INRIA team Odyssey, Brest, France}
\affiliation{4}{Institut Universitaire de France (IUF), Paris, France}
\affiliation{5}{Institut Pierre Simon Laplace, IRD, Sorbonne Université, Paris, France}




\correspondingauthor{Hugo Frezat}{hugo.frezat@univ-grenoble-alpes.fr}




\begin{keypoints}
\item Subgrid parametrizations can be learned end-to-end with a posteriori criteria involving model integration over several time-steps.
\item A posteriori learning for quasi-geostrophic turbulent flows can solve numerical stability issues related to energy backscatter.
\item Learned parametrizations outperform existing baselines for various evaluation metrics and apply to different flow configurations.
\end{keypoints}

%
%

%
%


\begin{abstract}
The use of machine learning to build subgrid parametrizations for climate models is receiving growing attention. State-of-the-art strategies address the problem as a supervised learning task and optimize algorithms that predict subgrid fluxes based on information from coarse resolution models. In practice, training data are generated from higher resolution numerical simulations transformed in order to mimic coarse resolution simulations. By essence, these strategies optimize subgrid parametrizations to meet so-called \textit{a priori} criteria. But the actual purpose of a subgrid parametrization is to obtain good performance in terms of \textit{a posteriori} metrics which imply computing entire model trajectories. In this paper, we focus on the representation of energy backscatter in two dimensional quasi-geostrophic turbulence and compare parametrizations obtained with different learning strategies at fixed computational complexity. We show that strategies based on \textit{a priori} criteria yield parametrizations that tend to be unstable in direct simulations and describe how subgrid parametrizations can alternatively be trained end-to-end in order to meet \textit{a posteriori} criteria. We illustrate that end-to-end learning strategies yield parametrizations that outperform known empirical and data-driven schemes in terms of performance, stability and ability to apply to different flow configurations. These results support the relevance of differentiable programming paradigms for climate models in the future.
\end{abstract}

\section*{Plain Language Summary} 
Climate projection and weather forecast heavily rely on computer simulations. 
But, if the physical laws governing the evolution of the climate system are well known, their simulation is still rather challenging. 
Fluid flows being essentially turbulent, small details at fine scales can have a tremendous impact on larger scales. 
Still, because of the limitations in computing power, all these interactions across scales cannot be explicitly resolved in computer simulations. 
Some of these interactions can only be represented approximately, and the design of these approximations is an active research area. 
We propose to train an approximate representation of unresolved scales of motions that optimizes the quality of the climate model over some temporal horizon, using machine learning. This results in more accurate and stable predictions.
Our method shows very promising results in toy example flow simulations, but its deployment at scale may seriously challenge the overall design of legacy climate models.

%
%

%


%
%
%
%

\section{Introduction}
The representation of unresolved processes is a key source of uncertainty in weather and climate models. Climate science and weather forecasting indeed heavily rely on numerical simulations of the Earth’s atmosphere and oceans  \cite{bauer2015quiet,neumann2019assessing}. But even the most advanced applications are currently far from resolving explicitly the wide variety of space-time scales and physical processes involved. This will likely remain the case for the foreseeable future because of the non-linearity of fluid dynamics and thermodynamics, and because of the finite nature of computational resources \cite{schneider2017climate,fox2014principles}. Weather and climate models will therefore keep relying on approximated representations of the effect of unresolved  processes in the form of subgrid \textit{parametrization} schemes \cite{schneider2017earth,fox2019challenges}. Parametrization schemes accounting for the impact of turbulence in the atmosphere and oceans at various scales will in particular remain essential components of these models. 

Parametrizations of unresolved turbulent motions are usually based on first principles, physics, idealized experiments, field observations and high resolution simulations. Their design involves a mixture of empirical and process-based modeling. Process-based models are formulated, tested and calibrated with experiments performed in the field, in the laboratory or with computers \cite{stensrud2009parameterization}. On this basis, the actual parametrization scheme estimates a tendency term for the target model from its resolved variables. The underlying conceptual framework can rely on some ensemble averaging procedure, so that the parametrization intends to capture the bulk statistical effect of unresolved processes, as for instance for turbulence models \cite{mellor1985ensemble}. Alternatively, one may consider a spatial filtering procedure and the parametrization can then exploit the scale-invariant properties of turbulence, as in Large Eddy Simulation (LES) models \cite{lesieur2005large}. This latter framework is used for instance for the parametrizations of ocean macro-turbulence in eddy-rich ocean models \cite{fox2008can}.  

Recently, the use of machine learning (ML) for better parametrizing unresolved processes in weather and climate models has gained momentum. Calibrating physics-based parametrization schemes against observation with ML and emulators is for instance becoming common practice \cite{ollinaho2013parameter,schneider2017earth,couvreux2021process}. Emulation approaches based on ML have also been proposed as a strategy for accelerating or regularizing existing schemes  \cite{ukkonen2020accelerating,chantry2021machine, meyer2021machine}. ML also provides new means to design new subgrid parametrization schemes from high fidelity simulations. In this context, ML may learn a mapping which predicts the tendency term due to unresolved subgrid effects from resolved quantities available in a target model. In atmospheric models, these approaches have been used to improve the representation of cloud micro-physics and moist processes \cite{krasnopolsky2013using,rasp2018deep,o2018using,brenowitz2018prognostic,seifert2020potential}. In ocean models, it is expected that the representation of macro-turbulence could be improved with similar approaches \cite{bolton2019applications,guillaumin2021stochastic}.

The design of parametrizations with ML builds on the rise of scientific machine learning (SciML) and its broad application to physical sciences. SciML is an emerging field, which bridges scientific computing and machine learning. Some recent key developments in this field have been motivated both \textit{from} physical insights and \textit{for} their applications to physical sciences, especially in fluid dynamics \cite{carleo2019machine,thuerey2021pbdl}. The conceptual developments in ML motivated by applications to problems governed by partial differential equations \cite{long2018pde,sirignano2018dgm,raissi2019physics} have for instance gradually freed ML from its black-box reputation. The design of parametrization schemes now directly benefits from ML approaches for dynamical system identification and equation discovery  \cite{brunton2016discovering,zanna2020data}. The ability to embed symmetries and law invariances into neural networks \cite{cohen2016group,cranmer2020lagrangian,alet2021noether} will also likely be important in the design of parametrization schemes \cite{frezat2021physical}, and in applications of ML to fluid mechanics in general \cite{brunton2020machine,vinuesa2021potential}. 

But ML-based approaches to subgrid parametrizations are still mostly based on \textit{a priori} learning strategies, which could limit their performance and applicability. There are indeed two different sorts of evaluation metrics for measuring the precision of subgrid models in turbulent simulations \cite{pope2000turbulent}. \textit{A priori} metrics, on the one hand, measure to what extent a given subgrid model is able to predict a tendency term due to unresolved subgrid effects at a fixed time. \textit{A posteriori} metrics, on the other hand, require to perform simulations with the subgrid model, and measure its integrated impact on the simulated flows. The common strategy for learning subgrid parametrizations is to formulate a supervised learning task from a high-resolution reference simulation dataset. In practice, learned parametrizations result from the minimization of a cost function based on some \textit{a priori} metrics measuring how well a mapping can predict unresolved fluxes from coarse-grain quantities. Still, with such strategy, what we \textit{really} intend to optimize is the ability of the parametrization to yield \textit{good} solutions, when used \textit{a posteriori} in numerical simulations. In principle, the versatility of ML algorithms should allow us to train parametrization schemes with learning criteria based on \textit{a posteriori} metrics, adopting the so-called \textit{end-to-end} learning framework \cite{glasmachers2017limits}. But surprisingly, there are very few published examples of end-to-end learning strategies in computational fluid dynamics \cite{sirignano2020dpm,kochkov2021machine,stachenfeld2021learned}. It is therefore yet unclear how subgrid parametrizations trained with \textit{a priori} and \textit{a posteriori} compare in terms of performance, stability and ability to apply to different flow conditions.

Flows governed by quasi-geostrophic (QG) dynamics provide an interesting and challenging testbed to evaluate learning strategies for subgrid parametrizations. 
Quasi-geostrophic theory indeed proposes a simple framework for studying geophysical flows constrained by earth rotation and stratification, 
as for instance large-scale atmospheric dynamics and ocean macro-turbulence \cite{cushman2011introduction}. 
The two-dimensional turbulence emerging from barotropic QG dynamics \cite{boffetta2012two,majda2006nonlinear} 
exhibits a dual cascade scenario with inverse energy transfers to larger scales and direct enstrophy transfers to smaller scales \cite{kraichnan1967inertial,thuburn2014cascades}. 
Because of this inverse energy cascade, developing subgrid parametrizations for QG flows is a challenging task, 
as the stability of numerical integration schemes is directly controled by the rate of energy backscatter, i.e. transfers from subgrid to resolved scales \cite{lilly1992proposed,carati1995representation}. 
As a consequence, a large number of subgrid parametrization schemes have been proposed for two-dimensional turbulence (see e.g. \citeA{danilov2019toward} for a review) and well documented flow configurations with performance metrics for parametrization are readily available \cite{graham2013framework}. 
Unsurprisingly, attempts to learn subgrid parametrizations for two-dimensional turbulence with ML have been less successful than for other types of turbulent flows \cite{maulik2019subgrid,guan2022stable}. 
In particular, ad-hoc solutions had to be implemented in order to ensure the numerical stability of the learned schemes under energy backscatter conditions. 
For instance, \citeA{maulik2019subgrid} uses a clipping post-processing procedure to remove negative diffusivity while \citeA{guan2022stable} mitigates this problem in decaying turbulence 
by increasing the size of the training dataset. 
More recently, \citeA{guan2022learning} and \citeA{pawar2022frame} demonstrated how incorporating physics in the models could lead to stable simulations that requires less data for training and generalizes better.
Up to now, however, none of the published works investigates the long-term statistical performance of learned schemes far beyond the decorrelation horizon. 
Learning stable parametrizations for two-dimensional turbulence in QG flows is therefore still an open problem. 

In this work, we compare parametrizations for two-dimensional turbulence obtained with different learning strategies, at fixed computational complexity. In particular, we show that we are able to train a model based on \textit{a posteriori} metrics with an end-to-end learning strategy. Through evaluation on three different configurations (decay, wind-forcing and beta-effect), the end-to-end learning strategy is shown to yield stable parametrizations that outperform previous physics-based and NN-based models without any explicit postprocessing step. Statistical metrics on long-term spectral transfers are shown to be in excellent agreement to direct numerical simulations (DNS), which is particularly encouraging for future climate models. The paper is organized as follows: In Sec. 2, we present the \textit{a priori} and \textit{a posteriori} learning strategies and the type of metrics they are respectively able to optimize. The application to quasi-geostrophic parametrizations is described in Sec. 3 with the numerical setup and baselines used in the evaluation. Results are presented both for short-term and long-term statistics for three different configurations in Sec. 4. Finally, we discuss the limitations and implications of the described strategies for realistic large-scale solvers.

\section{Learning strategies}

In this study, we address the simulation of the time evolution of geophysical quantities $\mathbf{y}(t)$. We assume the underlying governing equations to be known. Let us denote by 
$f(\mathbf{y})$ these true dynamics. The numerical integration of this system being either impossible or expensive, we aim at solving the time evolution of reduced variables $\bar{\mathbf{y}}(t)$ such that
\begin{align}
    \begin{cases}
    \dfrac{\partial \mathbf{y}}{\partial t} = f(\mathbf{y}), \hspace{5mm} &\mathbf{y} \in \Omega \\[.5ex]
    \dfrac{\partial \bar{\mathbf{y}}}{\partial t}  = g(\bar{\mathbf{y}}) + \mathcal{M}(\bar{\mathbf{y}}), \hspace{5mm} &\bar{\mathbf{y}} \in \bar{\Omega} \\[.5ex]
    \mathcal{T}(\mathbf{y}) = \bar{\mathbf{y}}\\
    \end{cases}
    \label{eq:reduced}
\end{align}
where $\bar{\Omega} \subset \Omega$, $g$ a reduced-order operator, $\mathcal{M}$ a subgrid-scale parametrization and $\mathcal{T}$ is a projection operator that maps true variables to reduced ones. The objective in reduced-order modeling is to design operator $g$ such that the evolution of the reduced variables match the projection $\mathcal{T}(\mathbf{y})$ of the true variables $\mathbf{y}$. We note that for some reduced order problems, we identify $f = g$ with variables existing on different spaces or dimensionalities.

Within a learning framework, one states the identification of subgrid-scale term $R(\mathbf{y}) = \mathcal{T}(f(\mathbf{y})) - g(\mathcal{T}(\mathbf{y}))$ as a learning problem from reduced variables for a parametrization $\mathcal{M}(\bar{\mathbf{y}} | \theta)$ where $\theta$ are trainable model parameters. Under the assumption that projection operator $\mathcal{T}$ commutes with partial derivatives, the most classic approach comes to train parametrization $\mathcal{M}(\bar{\mathbf{y}} | \theta)$ as a functional approximation of closure term $R$. This approach has been widely explored in the recent literature \cite{vollant2017subgrid,bolton2019applications}. 
It does not however constrain the trained parametrization to behave as expected when implemented in the solver of the reduced-order system. In this respect, an end-to-end framework would appear as an appealing approach to explicitly state the subgrid-scale parametrization problem according to the best possible approximation of the true reduced variables. Such end-to-end approaches have shown many advantages in the approximation of differential equation in general \cite{chen2018neural,bakarji2021data,fablet2021learning}. 
When applied to physical problems, they are often referred as differentiable physics \cite{de2018end,um2020solver,holl2020learning}, since they require the gradient of all the considered operators and solvers to be available for the optimization algorithm. Overall, these two categories of learning approaches differ in the space where the training is performed, similarly to the definition of \textit{a priori} and \textit{a posteriori} metrics \cite{pope2000turbulent} for the benchmarking of SGS parametrizations. This is the reason why we refer to \textit{a priori} and \textit{a posteriori} learning strategies as detailed in the subsequent.

\subsection{\textit{a priori} learning}
The \textit{a priori} learning strategy comes to learn SGS parametrization using training metrics defined on instantaneous quantities, i.e. a direct measure of the accuracy of the model based on the predicted SGS term $R(\mathbf{y})$. The \textit{a priori} loss $\mathcal{L}_{\mathrm{prio}}$ has the form,
\begin{equation}
    \mathcal{L}_{\mathrm{prio}}(\mathcal{M}) := \ell(R(\mathbf{y}), \mathcal{M}(\bar{\mathbf{y}} | \theta))
\end{equation}
where $\mathcal{M}$ is a given SGS model to be evaluated. The most common \textit{a priori} metrics $\ell$ found in the fluid dynamics community are the mean squared error (MSE) and the correlation between true and predicted SGS terms. Training a NN-based parametrization according to a \textit{a priori} setting then comes to building a representative ground-truth dataset  $\{R(\mathbf{y}_i),\bar{\mathbf{y}}_i\}_n$ of paired SGS terms and reduced variables to solve the following minimization problem w.r.t. model parameters $\theta$
\begin{equation}
    \operatorname*{arg\,min}_\theta \mathcal{L}_{\mathrm{prio}}(\mathcal{M}) \equiv \operatorname*{arg\,min}_\theta \ell(\{R(\mathbf{y}_i)\}, \mathcal{M}(\{\bar{\mathbf{y}}_i\} | \theta)).
    \label{eq:apriori}
\end{equation}
Solving for \eqref{eq:apriori} requires evaluation of the partial derivative of the \textit{a priori} loss $\mathcal{L}_{\mathrm{prio}}$ with respect to parameter $\theta$, which only involves the gradient of $\mathcal{M}$,
\begin{equation}
    \frac{\partial \mathcal{L}_{\mathrm{prio}}}{\partial \theta} = \frac{\partial \mathcal{L}_{\mathrm{prio}}}{\partial R} \frac{\partial R}{\partial \theta} + \frac{\partial \mathcal{L}_{\mathrm{prio}}}{\partial \mathcal{M}} \frac{\partial \mathcal{M}}{\partial \theta} = \frac{\partial \mathcal{L}_{\mathrm{prio}}}{\partial \mathcal{M}} \frac{\partial \mathcal{M}}{\partial \theta}.
\end{equation}
This approach has been applied to scalar \cite{vollant2017subgrid,portwood2021interpreting,frezat2021physical} and momentum \cite{gamahara2017searching,beck2019deep,xie2020modeling,yuan2020deconvolutional} parametrizations of three-dimensional turbulence on different configurations. The two-dimensional case is also well documented in decaying \cite{maulik2019subgrid,pawar2020priori,guan2022stable} and double-gyre \cite{bolton2019applications,zanna2020data} configurations. We may emphasize that, by construction, the \textit{a priori} learning strategy shall lead to the best \textit{a priori} results, which shall translate in a good instantaneous prediction of the SGS term according to metrics $\mathcal{L}_{\mathrm{prio}}$.

\subsection{\textit{a posteriori} learning}
\begin{figure}
    \centering
    \includegraphics{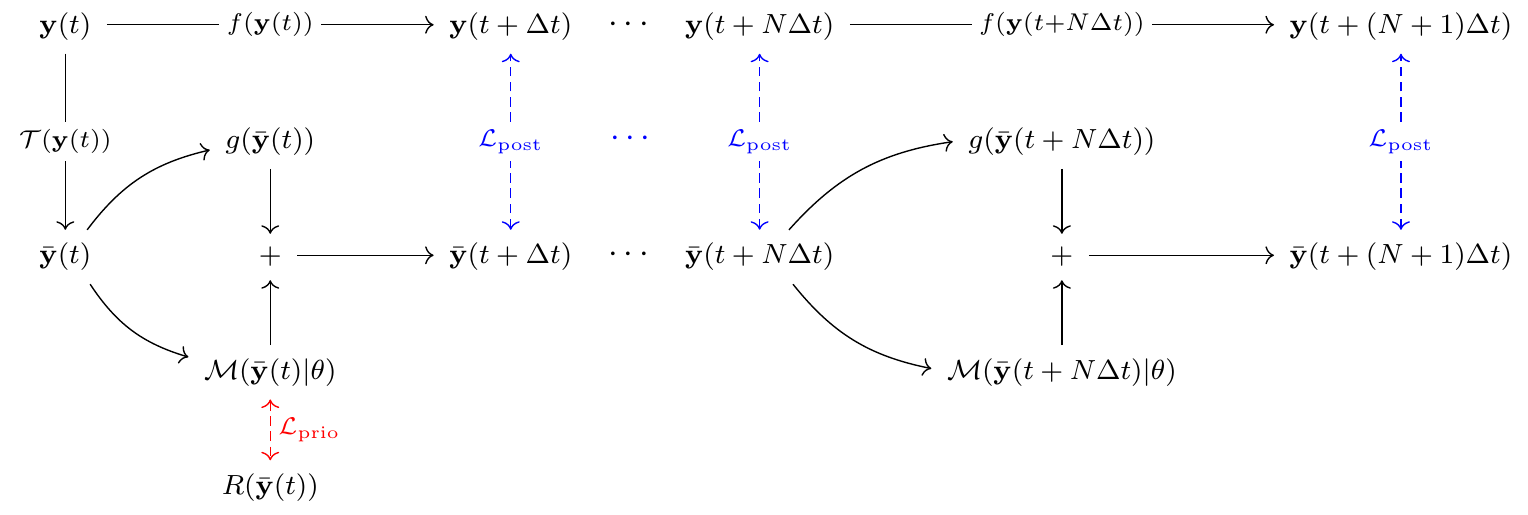}
    \caption{Sketch of one learning step for the \textit{a priori} and \textit{a posteriori} strategies. The \textit{a priori} loss is computed at instantaneous time $t$ (dashed, red), while the \textit{a posteriori} loss involves  several states forward in time (dashed, blue).}
    \label{fig:sketch_learning}
\end{figure}

The \textit{a posteriori} learning strategy states the SGS parametrization problem as the approximation of the true reduced variables according to some \textit{a posteriori} metrics. 
This is important since it is possible for a model to perform well \textit{a priori} while failing \textit{a posteriori},
the most common factor being numerical instabilities due to the lack of small-scale energy dissipation \cite{maulik2019subgrid,guan2022stable}. 
Let us denote by $\Phi$ the flow operator that advances the reduced system in time, i.e.,
\begin{equation}
    \Phi_{\theta}^{t_{1}}(\bar{\mathbf{y}}(t_{0})) = \bar{\mathbf{y}}(t_{0}) + \int_{t_{0}}^{t_{1}} g(\bar{\mathbf{y}}(t)) + \mathcal{M}(\bar{\mathbf{y}}(t) | \theta) \, \mathrm{d} t.
    \label{eq:flow}
\end{equation}
Numerically-speaking, flow operator $\Phi$ involves a time integration scheme (see Fig. \ref{fig:sketch_learning}) from start to end time $t_{0}$ and $t_{1}$, respectively. 
Following recent advances in neural integration schemes \cite{chen2018neural,ouala2021learning}, we may consider here both explicit and adaptive schemes. 
Let $\ell(\mathbf{y}(t), \bar{\mathbf{y}}(t))$ be some \textit{a posteriori} metrics defined in the true $\left\{ \mathbf{y}(t) \right\}_{t \in [0,T]}$ and 
reduced $\left\{ \bar{\mathbf{y}}(t) \right\}_{t \in [0,T]}$ spaces and a representative dataset of trajectories over a time interval $[0,T]$. 
The \textit{a posteriori} learning strategy comes to minimize an \textit{a posteriori} training loss,
\begin{equation}
    \mathcal{L}_{\mathrm{post}} := \ell (\left\{ \mathbf{y}(t) \right\}_{t \in [t_{0},t_{1}]}, \left\{ \Phi_{\theta}^{t}(\bar{\mathbf{y}}(t_{0})) \right\}_{t \in [t_{0},t_{1}]})
\end{equation}
Now, the \textit{a posteriori} minimization problem involves the time integration of $\Phi$ on sub-intervals $[t_{0},t_{1}]$ from the dataset of trajectories spanning the interval $[0,T]$, with initial reduced states typically taken to be $\bar{\mathbf{y}}(t_{0}) = \mathcal{T}(\mathbf{y}(t_{0}))$,
\begin{equation}
    \operatorname*{arg\,min}_\theta \mathcal{L}_{\mathrm{post}} \equiv \operatorname*{arg\,min}_\theta \ell(\left\{ \mathbf{y}(t) \right\}_{t \in [t_{0},t_{1}]}, \left\{ \Phi_{\theta}^{t}(\bar{\mathbf{y}}(t_{0})) \right\}_{t \in [t_{0},t_{1}]}), \forall [t_{0},t_{1}] \in [0,T].
    \label{eq:aposteriori}
\end{equation}
Then from the Leibniz integral rule, updating model parameters $\theta$ requires the flow partial derivative, i.e.
\begin{equation}
    \frac{\partial \mathcal{L}_{\mathrm{post}}}{\partial \theta} = \frac{\partial \mathcal{L}_{\mathrm{post}}}{\partial \mathbf{y}(t_{1})} \frac{\partial \mathbf{y}(t_{1})}{\partial \theta} + \frac{\partial \mathcal{L}_{\mathrm{post}}}{\partial \Phi} \frac{\partial \Phi}{\partial \theta} = \frac{\partial \mathcal{L}_{\mathrm{post}}}{\partial \Phi} \left( \bar{\mathbf{y}}(t_{0}) + \int_{t_{0}}^{t_{1}} \frac{\partial g(\bar{\mathbf{y}}(t))}{\partial \theta} + \frac{\partial \mathcal{M}(\bar{\mathbf{y}}(t) | \theta)}{\partial \theta} \, \mathrm{d} t \right).
\end{equation}
This equation makes explicit that the gradient-based optimization of the \textit{a posteriori} criterion involves the computation of the gradient with respect to all the components of the forward model, {\em i.e.} dynamical operator $g$ as well as the considered time integration scheme. Assuming that one can run all components within a differentiable programming framework (here, PyTorch \cite{paszke2019pytorch}), the embedded automatic differentiation tools makes these computations direct with no additional programming cost.
In our experiments, this comes to performing an automatic differentiation for an explicit fourth-order Runge-Kutta scheme with $N$ discrete time-steps, which defines the temporal horizon $T = N \Delta t$.

The \textit{a posteriori} strategy significantly widens the range of metrics which can be considered to calibrate the SGS parametrization. We illustrate this modeling flexibility for quasi-geostrophic turbulence in the next section. We may point out that this \textit{a posteriori} learning strategy has recently been explored for temporally-developing plane turbulent jets \cite{macart2021embedded} and the short-term simulation of short-term two-dimensional flows \cite{kochkov2021machine}. Here, we explore further its relevance for two-dimensional geophysical flows, including a benchmarking with the \textit{a priori} setting for different flow configurations.

\section{Application to quasi-geostrophic turbulence}
Geophysical turbulence is widely acknowledged to involve energy backscatter. This makes SGS parametrization a key issue for the simulation of ocean and atmosphere dynamics \cite{graham2013framework,jansen2015energy,juricke2020ocean}. As case-study framework, we consider quasi-geostrophic (QG) flows. While providing an approximate yet representative model for rotating stratified flows found in atmosphere and ocean dynamics, they involve relatively complex SGS features that make the learning problem non trivial. As such, QG flows are regarded as an ideal playground to explore and assess the relevance of \textit{a priori} and \textit{a posteriori} learning strategies for SGS parametrization in geophysical turbulence.
QG equations \cite{majda2006nonlinear} are given by,
\begin{equation}
    \partial_{t}\omega + J(\psi, \omega) = \nu \nabla^{2} \omega - \mu \omega - \beta \partial_{x} \psi + F
    \label{eq:qg}
\end{equation}
which is equivalent to the transport of vorticity $\omega$, obtained by taking the curl of the incompressible Navier-Stokes equations, i.e. $\nabla \cdot \mathbf{u} = 0$ and $\omega = \nabla \times \mathbf{u}$ and applying beta-plane approximation, hydrostatic and geostrophic balances. In addition, we have,
\begin{align}
    \mathbf{u} &= (-\partial_{y}\psi, \partial_{x}\psi)\\
    \omega &= \nabla^{2} \psi
\end{align}
where $\psi$ is the streamfunction, $\mathbf{u}$ the velocity and $J(\psi, \omega) = \partial_{x}\psi \partial_{y}\omega - \partial_{y}\psi \partial_{x}\omega$ is the non-linear Jacobian operator. The model is parametrized by viscosity $\nu$, linear drag coefficient $\mu$, Rossby parameter $\beta$ and a source term $F$. The QG equations have two invariants in the limit of inviscid flow \cite{bouchet2012statistical}, for energy
\begin{equation}
    E = \frac{1}{2} \int \mathbf{u}^2 \mathrm{d}r 
    \label{eq:energy}
\end{equation}
and enstrophy
\begin{equation}
    Z = \frac{1}{2} \int \omega^2 \mathrm{d}r.
    \label{eq:enstrophy}
\end{equation}
In order to study scales interactions, we introduce the enstrophy spectrum, $Z(k)$ in spectral space as the enstrophy contained in the shell of radius $k$,
\begin{equation}
Z(k) = \frac{1}{2}\int_{|\mathbf{k}|=k} \hat{\omega}(\mathbf{k}) \hat{\omega}^*(\mathbf{k}) \mathrm{d}S(\mathbf{k})
\end{equation}
where the Fourier transform is represented by $\hat{\cdot}$, complex conjugation by $\cdot^{*}$ and with $\mathrm{d}S(\mathbf{k})$ the integration element of the shell of radius $k$ from wavevector $\mathbf{k}$. The evolution equation of $Z(k)$ writes in spectral space,
\begin{equation}
\frac{d}{dt}Z(k) = -D(k) - Q(k) -B(k) + F(k) + T(k)
\end{equation}
where the different terms of the right hand side are related to various effects: external energy source $F$, dissipation $D$, large-scale drag $Q$, beta-plane effect $B$ and transfer between scales $T$. This last term writes
\begin{equation}
T(k) = \int_{|\mathbf{k}|=k}  \Re \{ \hat{\omega}^*(\mathbf{k})  \hat{J}(\psi, \omega)(\mathbf{k}) \} \mathrm{d}S(\mathbf{k}). 
\end{equation}
This allows to define the enstrophy flux \cite{gupta2019energy} through the wavenumber $k$ as 
\begin{equation}
\Pi_{Z}(k) = -\int_{0}^{k} T(k') \mathrm{d}k'
\label{eq:flux}
\end{equation}
which is a key quantity to measure effect of SGS modeling on enstrophy distribution on the range of resolved scales. 

\subsection{SGS parametrization for QG dynamics}
The derivation of the reduced model for QG dynamics follows the same procedure that is described for fluid dynamics in general. 
Assuming a known projection operator $\mathcal{T}$ from Eq. \eqref{eq:reduced} at spatial coordinate $\mathbf{x}$ given as a discretization 
$\mathcal{D}: \Omega \rightarrow \bar{\Omega}$ and the convolution of $\mathbf{y}$ with a kernel function $G(\mathbf{x})$ \cite{leonard1975energy},
\begin{equation}
    \bar{\mathbf{y}}(\mathbf{x}) = \mathcal{T}(\mathbf{y}(\mathbf{x})) = \mathcal{D} \left[ \int G(\mathbf{x} - \mathbf{x}') \mathbf{y}(\mathbf{x}') \mathrm{d} \mathbf{x}' \right].
    \label{eq:filtering}
\end{equation}
We can then derive the equations which govern the evolution of reduced vorticity $\bar{\omega}$ as,
\begin{align}
    \begin{cases}
    \partial_{t} \omega + J(\psi, \omega) = \nu \nabla^{2} \omega - \mu \omega - \beta \partial_{x} \psi + F, \hspace{5mm} &\omega \in \Omega \\[2ex]
    \partial_{t} \bar{\omega} + J(\bar{\psi}, \bar{\omega}) = \nu \nabla^{2} \bar{\omega} - \mu \bar{\omega} - \beta \partial_{x} \bar{\psi} + \bar{F} + \underbrace{J(\bar{\psi}, \bar{\omega}) - \overline{J(\psi, \omega)}}_{R(\psi, \omega)}, \hspace{5mm} &\bar{\omega} \in \bar{\Omega}
    \end{cases}
    \label{eq:qg_reduced}
\end{align}
where $R(\psi, \omega)$ is the SGS term. For convenience, note that the reduced term can be expressed in a flux formulation, 
\begin{equation}
    R(\psi, \omega) = \nabla \cdot \left( \bar{\mathbf{u}} \, \bar{\omega} - \overline{\mathbf{u} \, \omega} \right).
    \label{eq:rdivform}
\end{equation}
In this context, $\bar{\omega}$ is only solved for the largest scales of the flow, and $R(\psi, \omega)$ accounts for the effect of unresolved motions on the resolved scales. The SGS term is thus not known from the reduced variables because of the non-linear interactions of small-scale dynamics $\overline{J(\psi, \omega)}$. Following the notations introduced in Section 2, we aim to identify a QG SGS parametrization $\mathcal{M}(\bar{\psi}, \bar{\omega} | \theta)$ given the parametrization for operator $g$ by \eqref{eq:qg_reduced} using both \textit{a priori} and \textit{a posteriori} learning strategies. 

In order to study the effect of the SGS parametrization on scales interactions, we will consider the enstrophy spectrum $Z(k)$ and the associated enstrophy flux $\Pi_{Z}(k)$ in various flow configuration. Note that when the governing equation of reduced vorticity $\bar{\omega}$ is solved, the transfer term is now split in a resolved and a modeled part, as
\begin{equation} 
 T(k) = \int_{|\mathbf{k}|=k}  \Re \{  
 \underbrace{\hat{\bar{\omega}}^*(\mathbf{k})  \hat{J}(\bar{\psi}, \bar{\omega})(\mathbf{k})}_{\mathrm{resolved}} -
 \underbrace{\hat{\bar{\omega}}^*(\mathbf{k})  \hat{R}(\psi, \omega)(\mathbf{k})}_{\mathrm{modeled}} 
 \} \mathrm{d}S(\mathbf{k}). 
 \end{equation}
We can also look at the resolved enstrophy equilibrium in physical space. The governing equation for $\bar{z} = \frac{1}{2} \bar{\omega}^2$ writes
\begin{equation}
\partial_{t} \bar{z} =  \nu \bar{\omega} \nabla^{2} \bar{\omega} - \mu \bar{\omega} \bar{\omega} - \beta \bar{\omega} \partial_{x} \bar{\psi} + \bar{\omega} \bar{F} - \bar{\omega} J(\bar{\psi}, \bar{\omega}) + \underbrace{\bar{\omega} R(\psi, \omega)}_{T_{\bar{z}}}
\end{equation}
where the last term $T_{\bar{z}}$ shows the direct effect of the SGS term on the resolved enstrophy equilibrium. From Eq. \eqref{eq:rdivform}, this last term can be expressed as,
\begin{equation}
 T_{\bar{z}} = \nabla \cdot \left( \bar{\omega} (\bar{\mathbf{u}} \, \bar{\omega} - \overline{\mathbf{u} \, \omega}) \right) - (\bar{\mathbf{u}} \, \bar{\omega} - \overline{\mathbf{u} \, \omega}) \cdot  \nabla \bar{\omega},
\end{equation}
with a first term related to diffusion, and a second term related to the transfer between resolved and unresolved scales: if $(\bar{\mathbf{u}} \, \bar{\omega} - \overline{\mathbf{u} \, \omega}) \cdot  \nabla \bar{\omega} > 0$, there is a direct transfer from resolved to unresolved scales (forwardscatter), but if $(\bar{\mathbf{u}} \, \bar{\omega} - \overline{\mathbf{u} \, \omega}) \cdot  \nabla \bar{\omega} < 0$, there is a transfer from unresolved to resolved scale (backscatter). The latter will be the key quantity to study the ability of models to take into account backscatter. However, because models are built directly for $R(\psi, \omega)$ the above decomposition can not be performed, and only $T_{\bar{z}}$ can be considered.

\subsection{Numerical solver}
The equations \eqref{eq:qg} are solved using a pseudo-spectral differentiable code with full $3/2$ dealiasing \cite{canuto2012spectral} and a classical fourth-order Runge-Kutta time integration scheme. The system is defined in a squared domain $\Omega \in [-\pi, \pi]^2$ with a Fourier basis, i.e. double-periodic boundary conditions $\partial \Omega$ on $N_{\mathrm{true}} = (N_{x}, N_{y})$ grid points with uniform spacing $\Delta_{\mathrm{true}} = \Omega N_{\mathrm{true}}^{-1}$. The reduced states are obtained by projecting the true (or high-resolution) states through a convolution with a spatial kernel $G_{\delta}(k)$ at spatial scale $\delta > 0$ followed by a discretization on the reduced grid $\bar{\Omega}$, i.e. with larger spacing $\Delta_{\mathrm{reduced}} = \delta \Delta_{\mathrm{true}}$ equivalent to a sharp cutoff,
\begin{equation}
    \bar{\omega}(k) := (\omega * G_{\delta})(|k| < \pi \Delta_{\mathrm{reduced}}^{-1}).
\end{equation}
It has been shown previously that SGS parametrizations can perform differently depending on the type of filter used in the evaluations \cite{piomelli1988model,zhou2019subgrid}. We then aim to evaluate how the choice of the filter affects the learning strategies and we consider two common types of filters, defined in spectral space as
\begin{align}
    &\textit{Gaussian filter} : \nonumber\\
    G_{\delta}(k) &= \exp \left( - \frac{k^2 \Delta_{\mathrm{reduced}}^2}{24} \right),\\
    &\textit{Cut-off filter} :\nonumber\\
    G_{\delta}(k) &= 0, \,\, \forall k > \pi \Delta_{\mathrm{reduced}}^{-1}.
\end{align}
Regarding numerical aspects, we can solve the time integration of the reduced system $g$ with a larger time-step by a factor corresponding to the grid size ratio (or filter scale $\delta$), i.e. $\Delta t_{\mathrm{reduced}} = \delta \Delta t_{\mathrm{true}}$. To generate the corresponding datasets $\{R(\mathbf{y}_i),\bar{\mathbf{y}}_i\}_n$ and $\left\{ \mathbf{y}(t) \right\}_{t \in [0,T]}$, we subsample one true state every $\delta$ iterations performed by the true system $f$.

\subsection{Baseline parametrizations}
For benchmarking purposes, we implement some physics-based baselines and focus on parametrizations based on functional eddy viscosity \cite{kraichnan1976eddy}, i.e. models that artificially dissipate energy at relevant scales to remain stable. This is to be contrasted with structural models that produce backscatter and thus suffer from stability issues and will not be considered here. One can state these parameterizations in a flux formulation,
\begin{equation}
    \mathcal{M}_{\mathrm{P}}(\bar{\psi}, \bar{\omega}) = \nabla \cdot (\nu_{e} \nabla \bar{\omega}).
\end{equation}
where the eddy viscosity coefficient $\nu_{e}$ contains an arbitrary constant $c_{\mathrm{P}}$ for which the optimal value depends on the flow configuration,
\begin{equation}
    \nu_{e} = (c_{\mathrm{P}} \Delta)^{n} |\varepsilon_{\mathrm{P}}(\bar{\psi}, \bar{\omega})|
\end{equation}
with $n$ depending on the scaling law used to derive the model. 
We used the dynamic procedure proposed by \citeA{germano1991dynamic} and \citeA{lilly1992proposed} 
where the constant is computed from a least-square minimization of the residual SGS term with a filter size larger than $\delta$, 
i.e. $\tilde{\omega} = \omega * G_{\tilde{\delta}}, \tilde{\delta} > \delta$. 
We also apply spatial averaging with positive clipping \cite{pawar2020priori,guan2022stable} in order to avoid locally negative constants $c_{\mathrm{P}}(x, y) < 0$, i.e. ensuring that the models are purely diffusive and $\nu_{e} \geq 0$.

One of the most popular SGS model has been proposed by \citeA{smagorinsky1963general}. It derives from the assumption of direct cascade of energy, which is relevant for three-dimensional flows. However, this assumption is  expected not to translate well to two-dimensional or geophysical turbulence, even if it has been already employed in global climate models \cite{delworth2012simulated}. Following a similar derivation, the Leith model \cite{leith1996stochastic} is often referred as the two-dimensional counterpart of the Smagorinsky model, assuming a direct cascade of enstrophy. The models are defined as eddy viscosity coefficients proportional to the resolved strain rate $\bar{S}$ and vorticity gradient $\nabla \bar{\omega}$, respectively,
\begin{align}
    &\textit{Smagorinsky model} : \nonumber\\
    \nu_{e} &= (c_{\mathrm{S}} \Delta)^{2} |\bar{S}|,\\
    &\textit{Leith model} : \nonumber\\
    \nu_{e} &= (c_{\mathrm{L}} \Delta)^{3} |\nabla \bar{\omega}|.
\end{align}
We will denote by $\mathcal{M}_{\mathrm{DynSmagorinsky}}$ and $\mathcal{M}_{\mathrm{DynLeith}}$ the dynamic versions of these two models where $c_{\mathrm{P}}$ has been computed using the dynamic procedure mentioned above.

\subsection{Neural architecture and training}
Our main focus being here the impact of \textit{a priori} and \textit{a posteriori} strategies, we consider the same neural-network-based parametrization $\mathcal{M}$ with the two learning settings. We use a convolutional neural network (CNN) architecture, which is particularly relevant for translation-invariant problems and have been used with success to train SGS parametrizations, e.g. \cite{beck2019deep,bolton2019applications,lapeyre2019training,mohan2020embedding,frezat2021physical,guan2022stable}.

\begin{figure}
  \definecolor{conv}{HTML}{FFA500}
  \definecolor{relu}{HTML}{87CEFA}
  \definecolor{phys}{HTML}{9F2B68}
  \centering
    
    
    
  \includegraphics{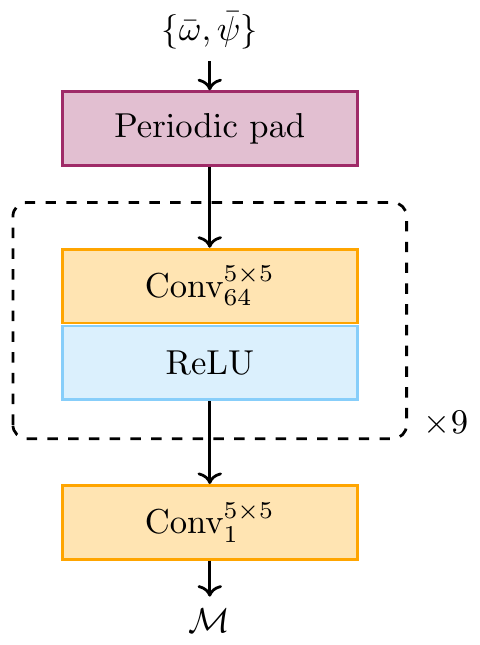}
  \caption{Sketch of the fully convolutional architecture employed in this study. The model applies 9 inner convolutional blocks, i.e. one 2D convolution layer $\text{Conv}_{C}^{K}$ and one non-linear ReLU activation. The number of channels $C$ for the inner convolutions and the kernel size $K$ are equal to $64$ and $5$, respectively. For the final layer, the model uses a 2D convolution layer to output the targeted real-valued SGS term. Since our computational domain is doubly periodic, we also use a periodic padding as a first layer.}
  \label{fig:architecture}
\end{figure}

As shown in Fig. \ref{fig:architecture}, we use a simple ConvNet with 10 layers of convolutions with non-linear ReLU activations. More involved architectures could further improve the performance. We may point out that our goal is not to design an optimal NN-based architecture, but rather to evaluate the impact of different learning strategies at similar computational complexity for the SGS parameterization.

Regarding the learning phase, the training loss for the \textit{a priori} strategy \eqref{eq:apriori} computes the MSE of the predicted term with respect to the true SGS term on a batch of $S$ samples,
\begin{equation}
    \mathcal{L}_{\mathrm{prio}}(\mathcal{M}) := \frac{1}{S} \sum_{s=1}^{S} || R(\psi, \omega)_s - \mathcal{M}(\bar{\psi_s}, \bar{\omega_s}) ||_{2}^{2}.
\end{equation}
For the \textit{a posteriori} strategy \eqref{eq:aposteriori}, the choice of training loss is more flexible, since we can explore spatio-temporal metrics for a batch of $N = T / \Delta t_{\mathrm{reduced}}$ discrete integration steps. To illustrate the basics of the strategy, we choose the MSE of the most important state of the QG system, i.e. the vorticity,
\begin{equation}
    \mathcal{L}_{\mathrm{post}}(\mathcal{M}) := \frac{1}{N} \sum_{n=1}^{N} ||\mathcal{T}(\omega(n \Delta t_{\mathrm{reduced}})) - \bar{\omega}(n \Delta t_{\mathrm{reduced}})||_{2}^{2}.
    \label{eq:post}
\end{equation}
The models are trained with the Adam optimizer on the same dataset containing 10 independent trajectories, i.e. simulations of 3000 snapshots each using different initial conditions, 
which gives a dataset of 30000 samples. As stated in \eqref{eq:post}, we subsample training data from the DNS sequences with a $\Delta t_{\mathrm{reduced}}$ sampling rate. We may point out that this configuration leads to a high inter-sample correlation for the training dataset \cite{guan2022learning}.
In our experiments, we chose an \textit{a priori} batch size equal to the number of \textit{a posteriori} temporal iterations, i.e. $S = N = 25$. This experimental setting ensures that both \textit{a priori} and \textit{a posteriori} training epochs use each training state only once within one epoch. It enables a fair comparison of the two schemes in terms of training convergence. Empirically, we noted that 30 epochs were necessary for the \textit{a priori} strategy, while 5 are enough for the \textit{a posteriori} strategy.
\begin{algorithm}
\caption{Training algorithm for SGS model $\mathcal{M}$ using the \textit{a posteriori} strategy. True variables $\mathbf{y}$ are sampled randomly from dataset which is only required to contain true variables. In practice, the projection can be applied beforehand and dataset $\left\{ \mathcal{T}(\mathbf{y}(t)) \right\}_{t \in [0,T]}$ can also be built from projected true states. Note that the outer loop iterates over the entire trajectories for each epoch.}
\label{alg:aposteriori}
\begin{algorithmic}[1]
\Require dataset $\left\{ \mathbf{y}(t) \right\}_{t \in [0,T]}$
\Require reduced system $g$, training model $\mathcal{M}(\bar{\mathbf{y}} | \theta)$
\Require number of iterations $N$, number of epochs $\epsilon$, starting time $t_{0}$
\Require loss function $\mathcal{L}_{\mathrm{post}}$
\For{$i \gets 1$ to $\epsilon$}
    \State $N_{\mathrm{eff}} \gets i \left \lfloor \frac{N}{\epsilon} \right \rfloor$ \Comment{Define effective number of iterations}
    \State $\left\{ \mathbf{y}(t) \right\} \gets \text{sample}(\left\{ \mathbf{y}(t) \right\}, N_{\mathrm{eff}})$ \Comment{Randomly sample consecutive true states}
    \State $\bar{\mathbf{y}}_{0} \gets \mathcal{T}(\mathbf{y}(t_{0}))$ \Comment{Define initial state from true states}
    \For{$j \gets 1$ to $N_{\mathrm{eff}}$}
        \State $\bar{\mathbf{y}}_{j} \gets \mathcal{I}(g(\bar{\mathbf{y}}_{j - 1}) + \mathcal{M}(\bar{\mathbf{y}}_{j - 1} | \theta))$ \Comment{Time integration with $\mathcal{I}$ (here RK4)}
    \EndFor
    \If{$\max_{j \in N_{\mathrm{eff}}} \mathrm{CFL}(\bar{\mathbf{y}}_{j}) > 1$} \Comment{Discard batch if stability is not obtained}
        \State continue
    \EndIf
    \State $\theta \gets \text{step}(\mathcal{M}, \frac{\partial \mathcal{L}_{\mathrm{post}}(\mathbf{y}, \bar{\mathbf{y}})}{\partial \theta})$ \Comment{Optimize model parameters from \textit{a posteriori} loss}
\EndFor
\end{algorithmic}
\end{algorithm}
We note however that the latter is consequently more expensive due to the solver steps involved in the training loop.
Regarding the \textit{a posteriori} strategy, training a model that performs the time integration of a system of PDEs inside the minimization loop may lead to instabilities and difficulties. 
To address these issues, we consider Algorithm \ref{alg:aposteriori}. It involves the following key steps:
\begin{itemize}
    \item A gradual increase of the time horizon $[0,T]$ (Line 2). The time integration scheme of reduced system with operators $g$ and $\mathcal{M}_{\mathrm{NN}}$ may result in a very deep computational graph (typically, ConvNet with more than 8 layers with the considered configuration with 25 integration steps), which may in turn lead to the commonly known vanishing gradient problem \cite{hochreiter2001gradient} especially for the first epochs of the training process. To address this issue, we gradually increase the temporal horizon from $[0,\frac{T}{\epsilon}]$ to $[0,T]$ where $\epsilon$ corresponds to the number of training epochs. In this study, the increment is done using a simple linear function, but any increasing heuristic should work as long as the first critical epochs take a small number of iterations.
    \item Withdrawing simulated data on the fly for which the Courant-Friedrichs-Lewy (CFL) is greater than some threshold (Line 8). Incorrect predictions from the NN especially during the first epochs of the training process for PDE problems at the limit of numerical stability can lead to numerical blowups of the system and by consequence exploding gradient for the minimization algorithm. We then discard batches for which the Courant-Friedrichs-Lewy (CFL) is greater than some threshold, commonly chosen to be 1.
\end{itemize}

\begin{figure}
    \centering
    \noindent\includegraphics[width=.8\textwidth]{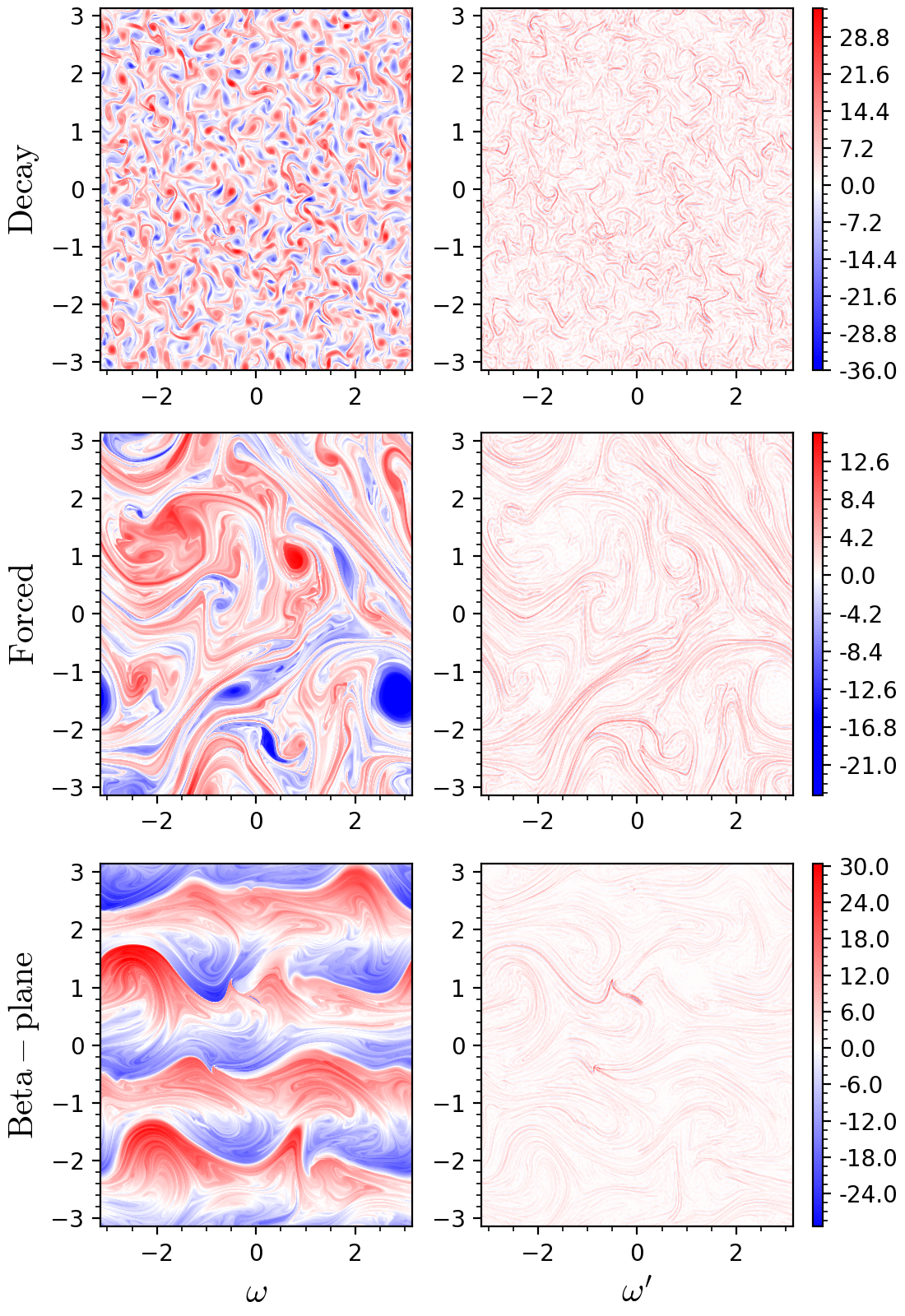}
    \caption{True initial vorticity field $\omega$ (left) and corresponding sub-grid contribution $\omega^{\prime} = \omega - \bar{\omega}$ (right) for the three case-studies: decay (top), forced (middle) and beta-plane (bottom).}
    \label{fig:dns_fields}
\end{figure}

\section{Results}
In order to evaluate the performance of \textit{a priori} and \textit{a posteriori} learning strategies, we report numerical experiments for three different configurations of QG flows (see \ref{fig:dns_fields}). First, we study decaying turbulence and compare the \textit{a posteriori} strategy with previous works based on the \textit{a priori} strategy \cite{maulik2019subgrid,guan2022stable}. Then, we assess the performance of the proposed models in a more realistic wind-forced configuration representative of mesoscale oceanic simulations \cite{fox2008can,graham2013framework}. Finally, we analyze the impact of planetary rotation through the beta-plane effect on a mid-latitude geophysical flow.

For these three cases, we consider the following experimental setup. The training and test data involve respectively 10 and 5 direct numerical simulations (DNS) corresponding to the same configurations with different initial conditions. The reduced systems are run with $\delta = 16$, i.e. the reduced grid is 16 times smaller compared to the true grid resolution in each direction. Reduced systems are integrated for 6000 iterations for the non-stationary decay cases and 18000 iterations to determine long-term statistics of the forced and beta-plane configurations. We may emphasize that the simulations used for evaluation purposes are never seen during the training phase. The parameters of the different flows are shown in Table \ref{tbl:parameters} in dimensionalized units. Overall, for each QG configuration, we report a quantitative synthesis for the cutoff and Gaussian filters and further illustrate the key features of the different learning strategies using a cutoff filter. 

\begin{table}
    \caption{Parameters of the different DNS flow configurations. 
    Details about simulation spin-up, initialization and forcing parameters are described in more details in Sec. 4.2. 
    Note that reduced systems use the same parameters, except for grid resolution ($\bar{N}_x, \bar{N}_{y}$) obtained from the spatial filter scale $(N_x / \delta, N_y / \delta)$ 
    and time-step $\Delta t_{\mathrm{reduced}} = \delta \Delta t_{\mathrm{true}}$. 
    The quantities are given as numerical (unitless) values directly used in the solver for reproductibility. Still, details on how these parameters are chosen are provided in their corresponding result section.}
    \label{tbl:parameters}
    \centering
    \begin{tabular}{l c c c c c c c c}
        \hline
         Name & $N_x \times N_y$ & $L_x \times L_y$ & $\Delta t$ & $\mu$ & $\nu$ & $\beta$ & Re \\
        \hline
        Decay       & $2048 \times 2048$ & $2 \pi \times 2 \pi$ & $10^{-4}$ & $0$                & $3.125 \times 10^{-5}$ & $0$                 & $3.2 \times 10^{4}$ \\
        Forced      & $2048 \times 2048$ & $2 \pi \times 2 \pi$ & $10^{-4}$ & $2 \times 10^{-2}$ & $1.025 \times 10^{-5}$ & $0$                 & $2.2 \times 10^{5}$ \\
        Beta-plane  & $2048 \times 2048$ & $2 \pi \times 2 \pi$ & $10^{-4}$ & $2 \times 10^{-2}$ & $1.025 \times 10^{-5}$ & $2.195 \times 10^{2}$ & $3.4 \times 10^{5}$ \\
        \hline
    \end{tabular}
\end{table}
Important quantities discussed in the evaluation are both;
\begin{itemize}
    \item \textbf{short term} temporal evolution of quadratic invariants both for the energy \eqref{eq:energy} and the enstrophy \eqref{eq:enstrophy}, typical in weather forecast.
    \item \textbf{long term} statistics from enstrophy spectrum and flux \eqref{eq:flux} in spectral space, relevant for climate predictions.
\end{itemize}

\subsection{Decaying turbulence}
In the context of two-dimensional SGS parametrization with ML models, the decaying turbulence configuration is one of the most studied \cite{maulik2019subgrid,pawar2020priori,guan2022stable}. 
This type of flow is particularly interesting because of its non-stationary nature, i.e. the system invariants are temporally varying. 
Similarly to \citeA{guan2022stable}, we sample the initial vorticity fields randomly from a gaussian distribution $\omega \sim \mathcal{N}(0, 1)$ at moderate wavenumbers $k \in [10, 32]$ 
and integrate the system for 10000 iterations before reaching spectrum self-similarity \cite{batchelor1969computation}. 

From this starting time $t_{0}$, the vorticity fields exhibit an early turbulence behavior with a lot of fine structures (see Fig. \ref{fig:dns_fields}, top). 
DNS and reduced models with SGS parametrizations are run until $t = 9.6$, which is longer than the temporal horizon used in the training data by a factor of two. 
In two-dimensional decaying flows, we expect to see vortex pairing and emergence of larger structures as shown in Fig. \ref{fig:eval_quantities} (left). 
Due to their purely diffusive form, parameterizations $\mathcal{M}_{\mathrm{DynSmagorinsky}}$ and $\mathcal{M}_{\mathrm{DynLeith}}$ cannot perform well on this configuration 
and are indeed incorrectly dissipating relevant small scales. 
The NN-based model trained with the \textit{a priori} strategy has accumulated small-scale enstrophy and is thus perturbed with noise coming from numerical instabilities. 
The model trained with the \textit{a posteriori} visually shows the expected stable dynamics with small-scale features, even outside the training regime, 
which supports some degree of generalization (or extrapolation).

\begin{figure}
    \centering
    \noindent\includegraphics[width=.8\textwidth]{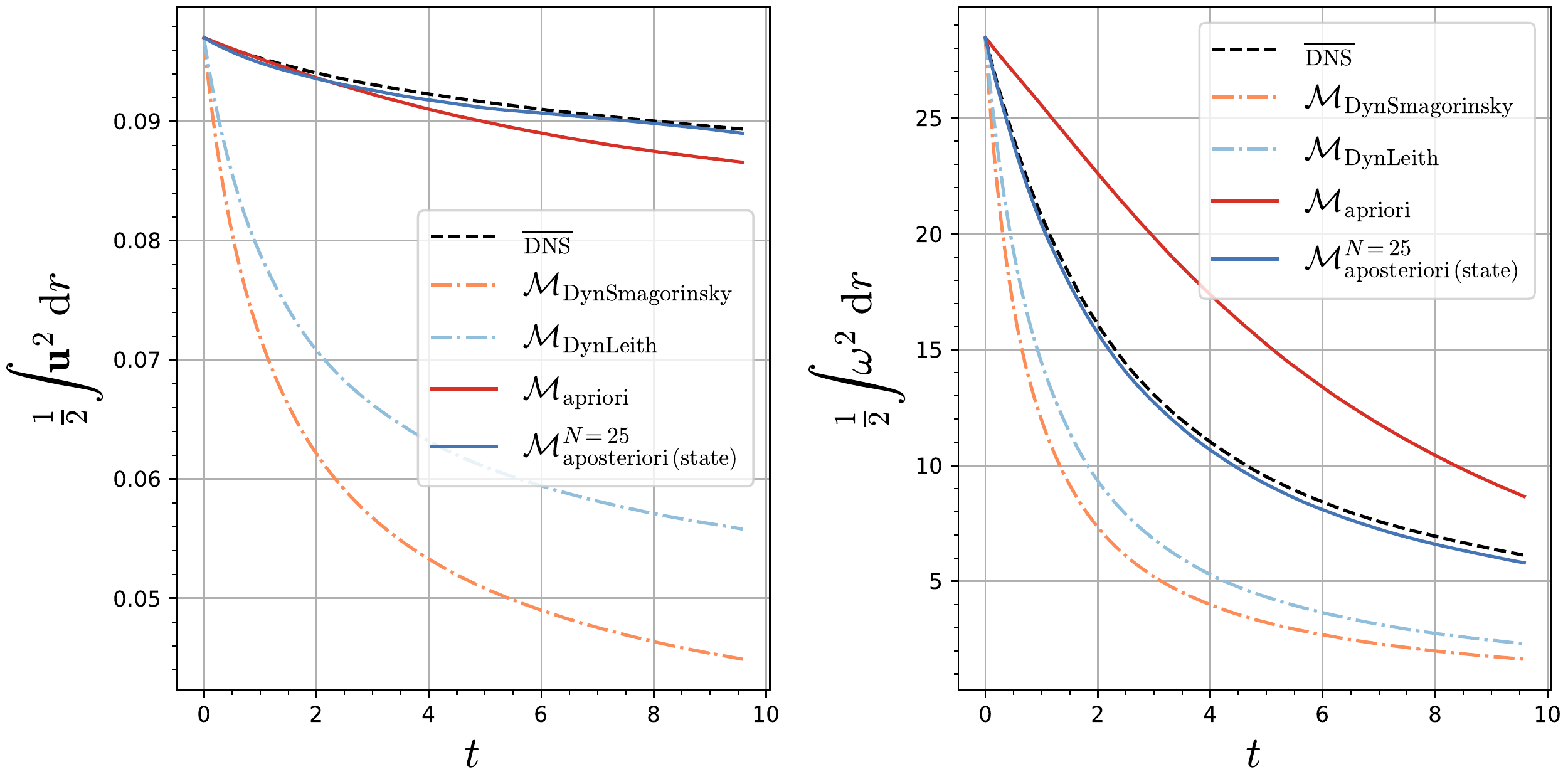}
    \caption{Evolution of domain-averaged energy (left) and enstrophy (right) computed in non-dimensionalized time units in the decaying turbulence setting.}
    \label{fig:decay_integrals}
\end{figure}

\begin{figure}
    \centering
    \noindent\includegraphics[width=.8\textwidth]{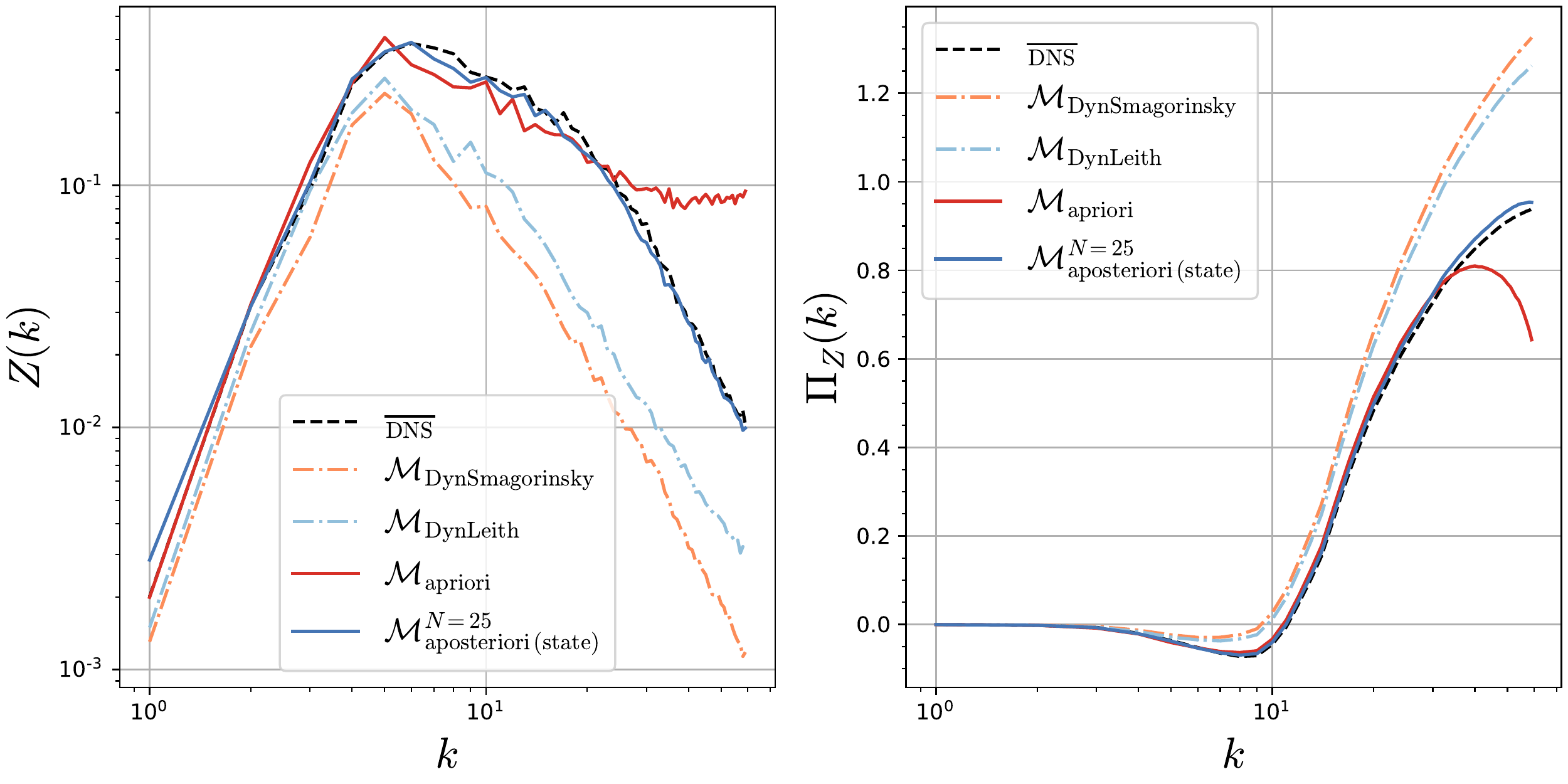}
    \caption{Final enstrophy spectrum (left) and time-averaged enstrophy flux (right) describing statistical performance of the models in decaying turbulence.}
    \label{fig:decay_stats}
\end{figure}

The evolution of domain-averaged quadratic integrals in Fig. \ref{fig:decay_integrals} confirms the observations from the vorticity fields, 
since we can see a large energy and enstrophy decrease for both $\mathcal{M}_{\mathrm{DynSmagorinsky}}$ and $\mathcal{M}_{\mathrm{DynLeith}}$, 
while the \textit{a priori} model correctly captures the energy decay but dissipates enstrophy too slowly compared to the DNS. 
Spectral statistics shown in Fig. \ref{fig:decay_stats} are also in close agreement to the DNS for the \textit{a posteriori}-trained model, 
in particular for the large wavenumbers of the enstrophy spectrum $Z(k) = k^2 E(k)$ which particularly highlight the dynamics of the smallest resolved scales.

\subsection{Forced turbulence}
The second case-study involves QG flows with a source term $F$ designed to mimic wind-stress. 
We study a particular configuration inspired by \citeA{graham2013framework}, which evaluated the performance of a large number of physics-based parametrizations in mesoscale ocean simulations. 
To reproduce these realistic equilibrium solutions, we use a bottom drag ($\mu > 0$) and initiate turbulent mixing from a wind-stress slowly varying in time at large-scale $k = 4$ 
with steady enstrophy rate injection $Z(F) = 3$ such that,
\begin{align}
F_{\omega}(t) &= \cos(4y + \pi \sin(1.4t))\\
	      &- \cos(4x + \pi \sin(1.5t))\nonumber\\
            F &= \frac{1}{3} Z(F_{\omega}) F_{\omega}(t)
\end{align}
In order to converge to a stationary turbulent state, we initialize the simulation runs from a few large-scale Fourier modes and spin-up on a smaller grid ($1024^2$) for over 500000 iterations. The initial conditions for training and evaluation (see Fig. \ref{fig:dns_fields}, middle) are taken after energy and enstrophy propagation to the smallest scales of the true grid ($2048^2$) in about 25000 iterations. 

To evaluate the long-term performance of the forced configuration in equilibrium, we run simulations until $t = 2.88$, which is at least 3 times longer than the complete decorrelation time of the system due to chaos. 
We report the vorticity fields at the end of the simulations in Fig. \ref{fig:eval_quantities} (center). 
The true vorticity state exhibits both large vortices generated by the wind forcing and small filaments in between. 
Overall, we draw conclusions similar to the decaying turbulence regime. 
We note that the small structures are inaccurately predicted for both $\mathcal{M}_{\mathrm{DynSmagorinsky}}$ and $\mathcal{M}_{\mathrm{DynLeith}}$ due to dissipation 
and for the \textit{a priori} model due to numerical instabilities. 
By contrast, the \textit{a posteriori} model is the only one to correctly capture both the large-scale and fine-scale patterns in this configuration.

\begin{figure}
    \centering
    \noindent\includegraphics[width=.8\textwidth]{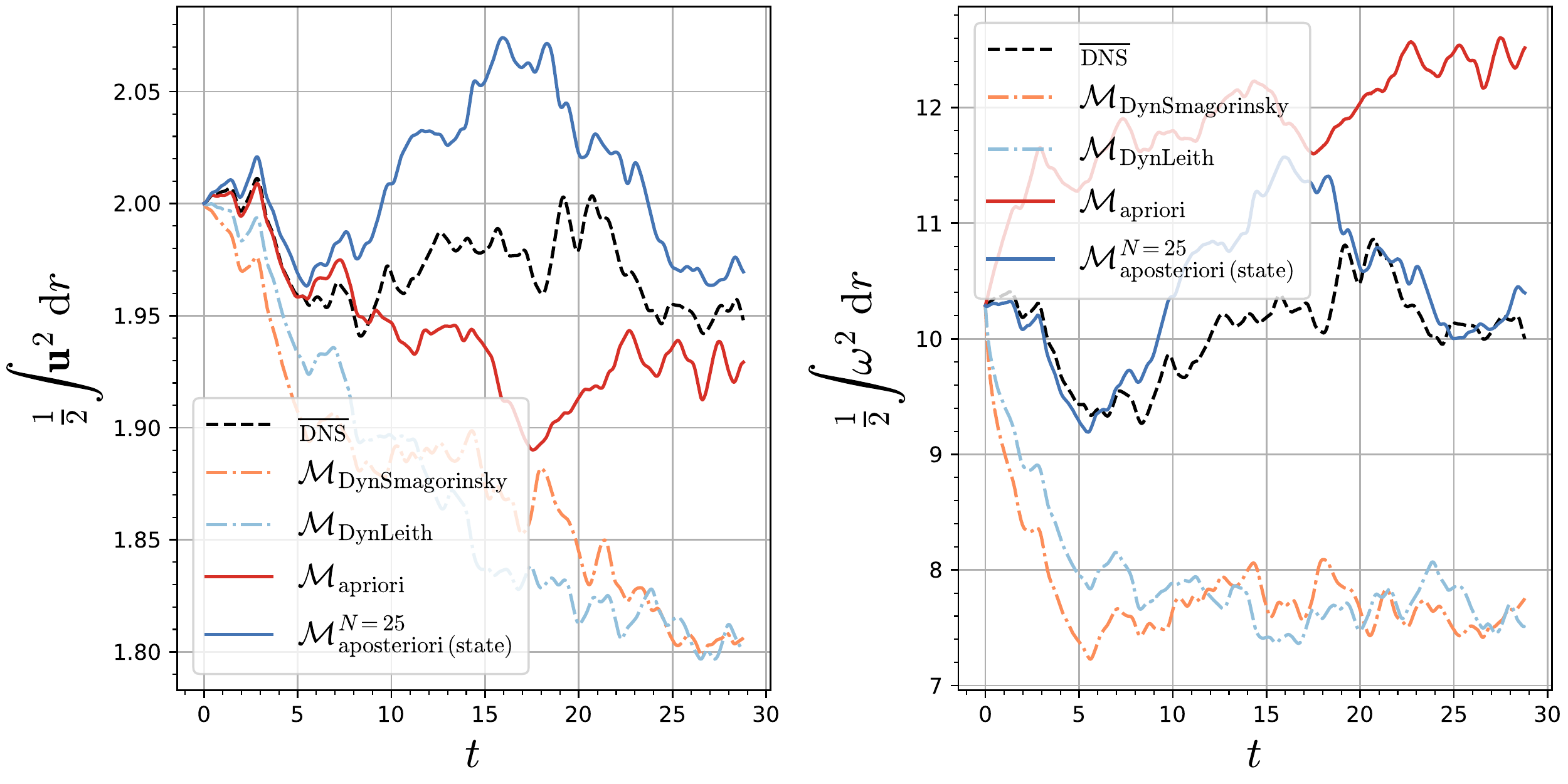}
    \caption{Evolution of domain-averaged energy (left) and enstrophy (right) computed in non-dimensionalized time units in the forced turbulence setting.}
    \label{fig:forced_integrals}
\end{figure}

\begin{figure}
    \centering
    \noindent\includegraphics[width=.8\textwidth]{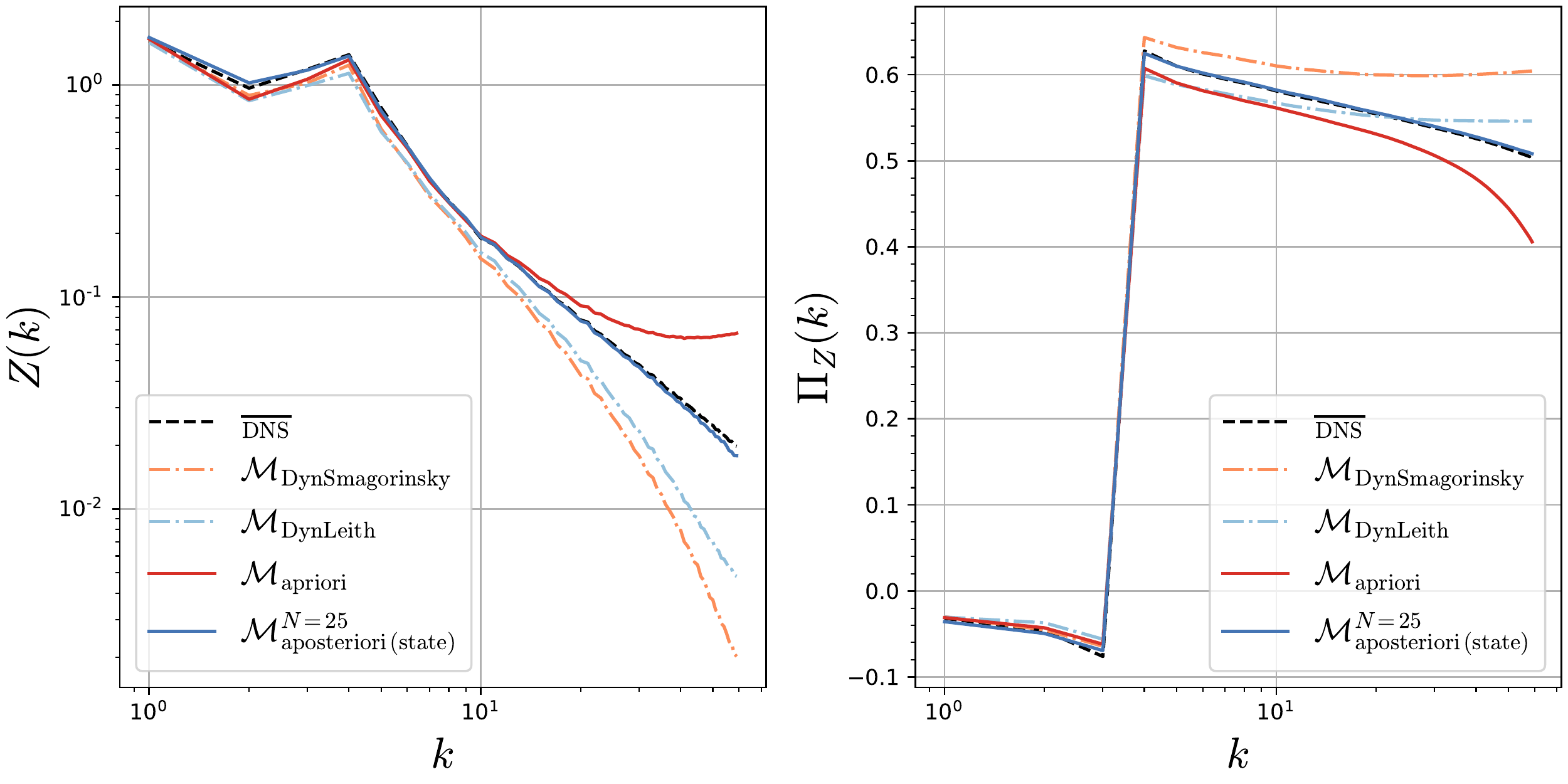}
    \caption{Time-averaged enstrophy spectrum (left) and enstrophy flux (right) describing statistical performance of the models in forced turbulence.}
    \label{fig:forced_stats}
\end{figure}

While domain-averaged integrals fluctuate a lot on such long-term trajectories due to the chaotic nature of the flow, we expect those quantities to remain approximately constant over time. 
This property is verified on the kinetic energy for both NN-based models, but the enstrophy of the \textit{a priori}-trained model increases over time, 
which indicates some accumulation of small-scale energy (see Fig. \ref{fig:forced_integrals}) and may result in a potential future blow-up of the simulation. 
However, the time-averaged statistical enstrophy spectrum shown in Fig. \ref{fig:forced_stats} demonstrates the ability of the \textit{a posteriori} model to reproduce accurately both 
the smallest scales and the largest scales of the simulation (small wavenumbers) compared to the other models. 

\subsection{Beta-plane turbulence}
The third case-study runs the same forced simulation as in the previous configuration complemented by a beta-plane effect to account for the meridional variation of Coriolis force caused by a spherical shape. We take a Rossby parameter $\beta$ corresponding to Earth's planetary rotation on mid latitudes (60°).

\begin{figure}
    \centering
    \noindent\includegraphics[width=.9\textwidth]{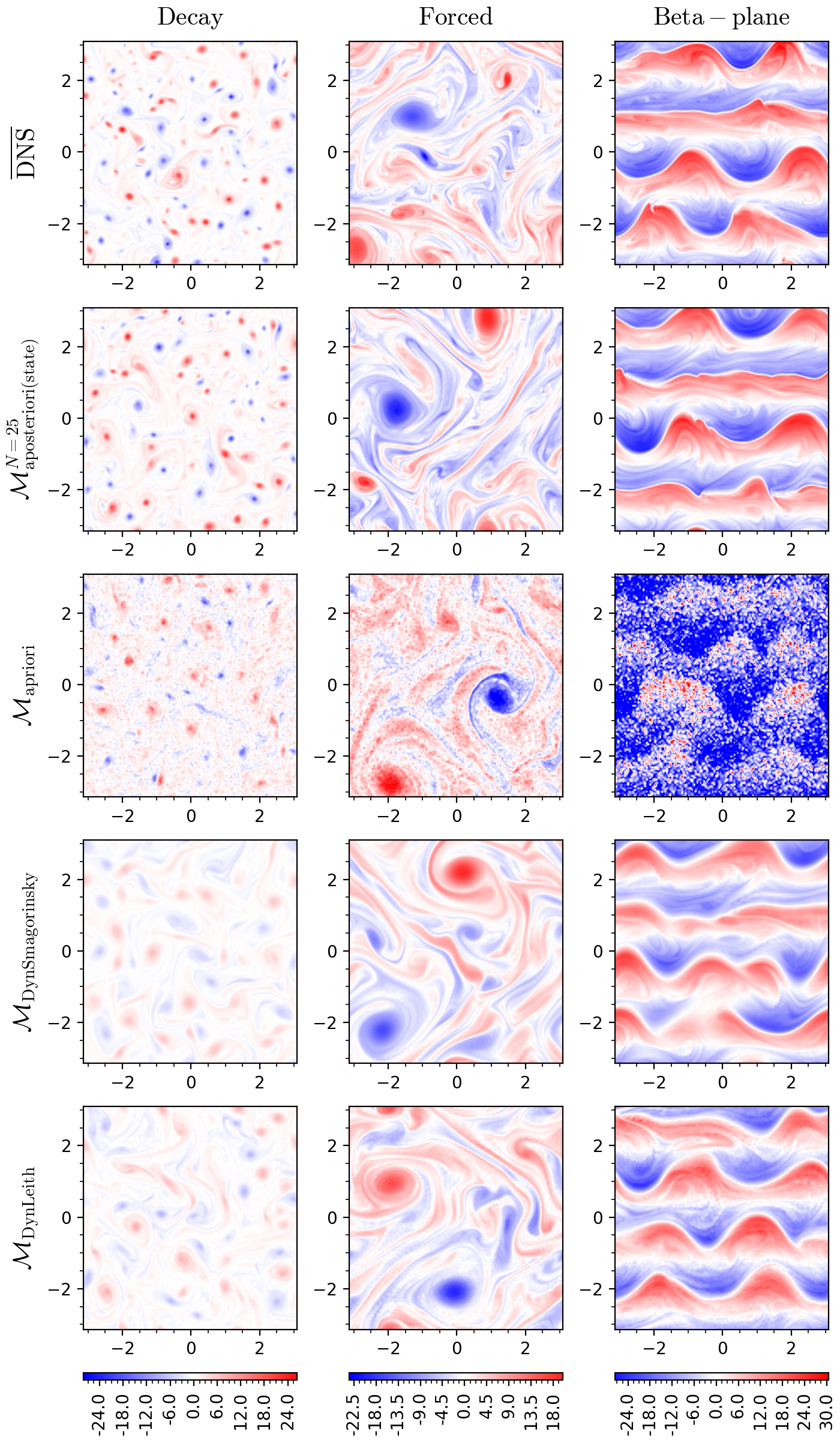}
    \caption{Vorticity fields for the different models at the end of one turbulence evaluation trajectory in each configuration.}
    \label{fig:eval_quantities}
\end{figure}

The beta-effect has an important impact on the topology of the dynamics, as it creates high-velocity longitudinal jets as seen in Fig. \ref{fig:dns_fields}, bottom. In this simple setting without topography (flat bottom layer), the system does not go through state transitions but remains in a statistical equilibrium. The strong vorticity gradients in between jets seen from \ref{fig:eval_quantities} (right) are predicted more accurately by the $\mathcal{M}_{\mathrm{DynLeith}}$ than the $\mathcal{M}_{\mathrm{DynSmagorinsky}}$, which still over-dissipates at small-scale. The model trained \textit{a priori} is not stable at all in this configuration, while the \textit{a posteriori} model remains stable to simulate visually-consistent patterns. 

\begin{figure}
    \centering
    \noindent\includegraphics[width=.8\textwidth]{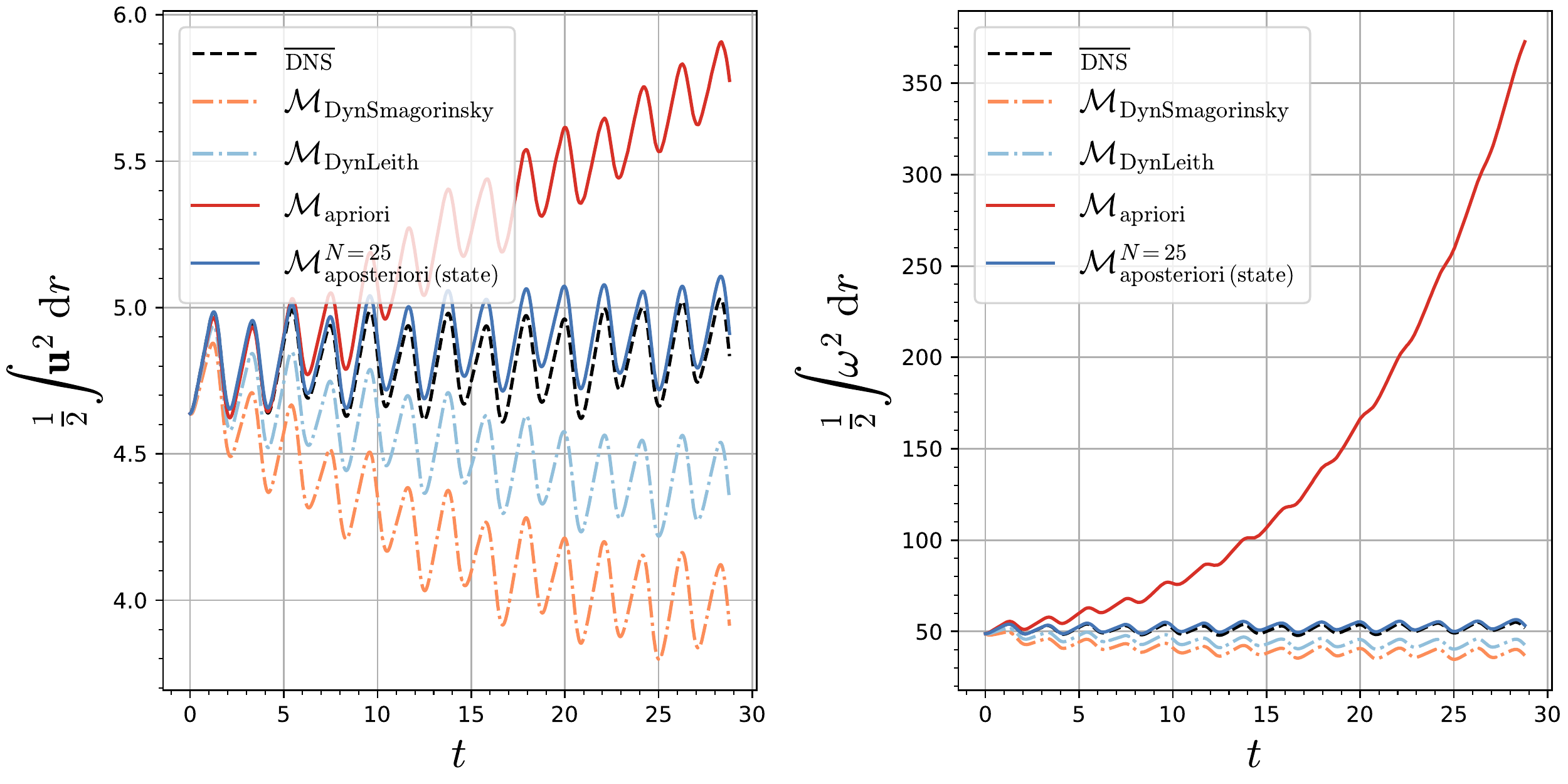}
    \caption{Evolution of domain-averaged energy (left) and enstrophy (right) computed in non-dimensionalized time units in the beta-plane turbulence setting.}
    \label{fig:geo_integrals}
\end{figure}

\begin{figure}
    \centering
    \noindent\includegraphics[width=.8\textwidth]{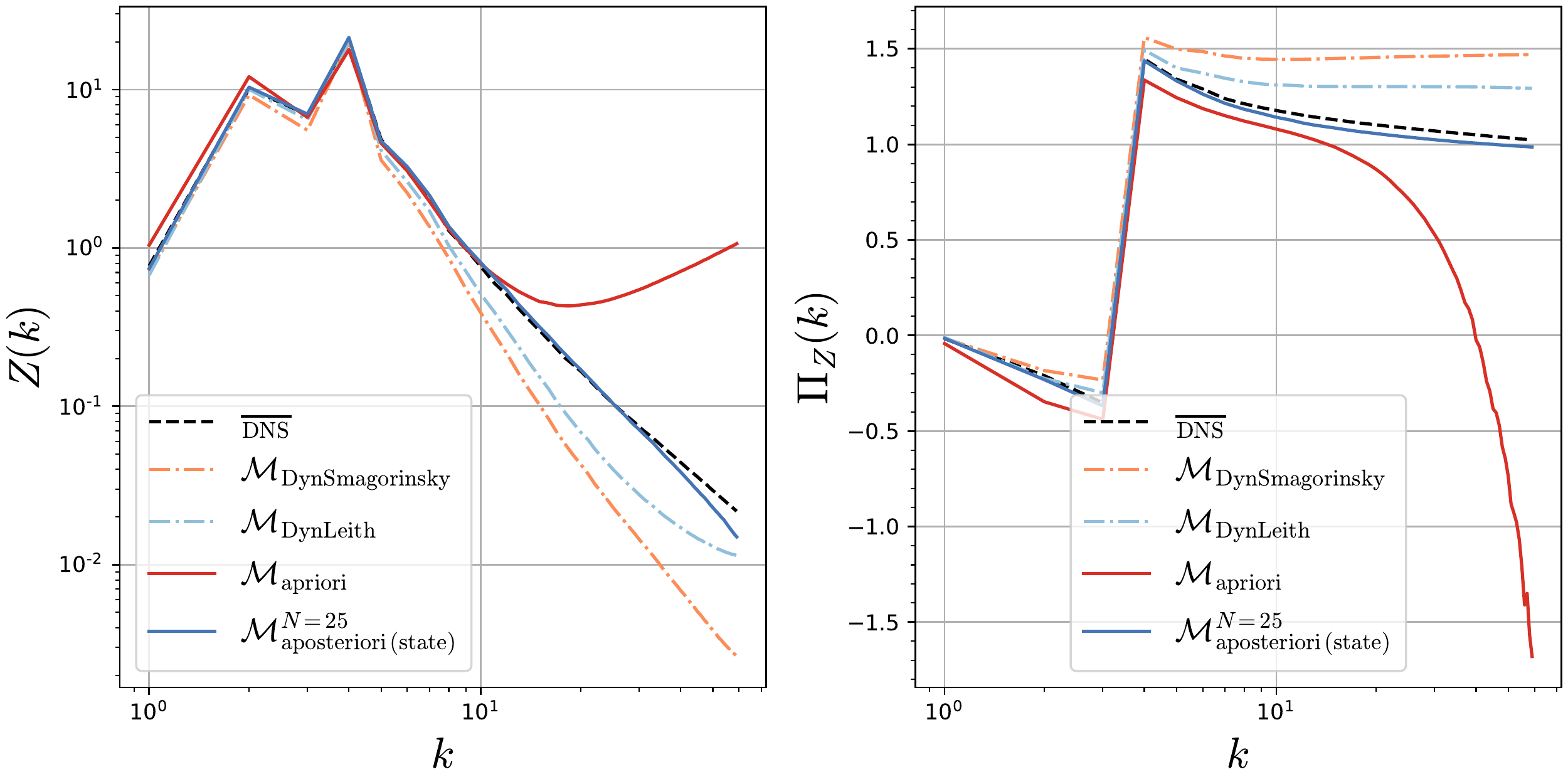}
    \caption{Time-averaged enstrophy spectrum (left) and enstrophy flux (right) describing statistical performance of the models in beta-plane turbulence.}
    \label{fig:geo_stats}
\end{figure}

The instabilities visible in the vorticity field of the \textit{a priori} model are explained by the increasing energy and enstrophy in Fig. \ref{fig:geo_integrals}. 
This reveals a non-conservative behavior which leads to a simulation blowup. 
The \textit{a posteriori} strategy performs extremely well on this long-term simulation, predicting in particular a correct enstrophy evolution compared to that of the DNS. 
The enstrophy spectrum and fluxes (see Fig. \ref{fig:geo_stats}) are similar to those of the forced turbulence configuration, 
except that the linear damping has a stronger impact, due to the relative increase in velocity from the beta-effect. 
Overall, the conclusions are the same as the previous two configurations, with highest fidelity small-scale dynamics being produced by the model trained using the \textit{a posteriori} strategy.

\subsection{Quantitative analysis}
As a quantitative synthesis of our numerical experiments, 
we first report the performance on two metrics for the three case-studies and the different models: an \textit{a priori} metric given by Pearson correlation coefficient of the predicted SGS terms (Table. \ref{tbl:apriori}.)
and an \textit{a posteriori} metric given by a variant of the error-landscape enstrophy flux assessment presented by \citeA{meyers2011error} (Table. \ref{tbl:aposteriori}). 
We report these performance metrics both for cutoff and Gaussian filters.

\begin{table}
    \caption{Short-term performance of the considered SGS parametrizations in the three different configurations with both cutoff and gaussian projection kernels. We compute the correlation coefficient between predicted and exact subgrid terms, which favors the \textit{a priori} learning strategy.}
    \label{tbl:apriori}
    \centering
    \begin{tabular}{l c c c c c c}
        \hline
         $\rho_{R, \, \mathcal{M}}$ & \multicolumn{2}{c}{Decay} & \multicolumn{2}{c}{Forced} & \multicolumn{2}{c}{Beta-plane} \\
         & Cutoff & Gaussian & Cutoff & Gaussian & Cutoff & Gaussian \\
        \hline
        $\mathcal{M}_{\mathrm{DynSmagorinsky}}$ & 0.16 & 0.38 & 0.09 & 0.55 & 0.04 & 0.28 \\
        $\mathcal{M}_{\mathrm{DynLeith}}$ & 0.13 & 0.32 & 0.08 & 0.49 & 0.03 & 0.17 \\
        $\mathcal{M}_{\mathrm{a priori}}$ & 0.75 & \textbf{0.90} & \textbf{0.82} & \textbf{0.95} & \textbf{0.82} & \textbf{0.96} \\
        $\mathcal{M}_{\mathrm{a posteriori \, (states)}}$ & \textbf{0.77} & 0.57 & 0.45 & 0.29 & 0.48 & 0.21\\
        \hline
    \end{tabular}
\end{table}

\begin{table}
    \caption{Long-term performance of considered SGS parametrizations in the three different configurations with both cutoff and gaussian projection kernels. We compute the $L^2$ distance between the reference enstrophy fluxes and the ones simulated using the different SGS parameterizations. The \textit{a posteriori} learning strategy clearly leads to much better scores.}
    \label{tbl:aposteriori}
    \centering
    \begin{tabular}{l c c c c c c}
        \hline
         $L^{2}(\Pi_{Z}^{R} - \Pi_{Z}^{\mathcal{M}})$ & \multicolumn{2}{c}{Decay} & \multicolumn{2}{c}{Forced} & \multicolumn{2}{c}{Beta-plane} \\
         & Cutoff & Gaussian & Cutoff & Gaussian & Cutoff & Gaussian \\
        \hline
        $\mathcal{M}_{\mathrm{DynSmagorinsky}}$ & 1.95 & 1.31 & 0.49 & 0.16 & 2.83 & 1.75\\
        $\mathcal{M}_{\mathrm{DynLeith}}$ & 1.64 & 1.02 & 0.16 & 0.11 & 1.66 & 0.98\\
        $\mathcal{M}_{\mathrm{a priori}}$ & 0.74 & 0.60 & 0.36 & 0.40 & 8.73 & 0.26\\
        $\mathcal{M}_{\mathrm{a posteriori \, (states)}}$ & \textbf{0.13} & \textbf{0.09} & \textbf{0.02} & \textbf{0.02} & \textbf{0.30} & \textbf{0.05}\\
        \hline
    \end{tabular}
\end{table}

The selected learning strategy clearly impacts the corresponding metrics. For the three flow regimes, 
the \textit{a priori} learning performs better on \textit{a priori} metrics, whereas  the \textit{a posteriori} learning leads to the best score on \textit{a posteriori} metrics.
Concerning the projection kernel, physical models and the NN-based model trained \textit{a priori} perform better with the Gaussian filter. Note that this is not surprising, 
as Gaussian filtering tends to smooth out discontinuities in the subgrid term \cite{canuto2007spectral} which might help during \textit{a priori} training. By contrast, the \textit{a posteriori} learning scheme results in consistent scores for both the Gaussian and cutoff kernels.

\begin{figure}
    \centering
    \noindent\includegraphics[width=\textwidth]{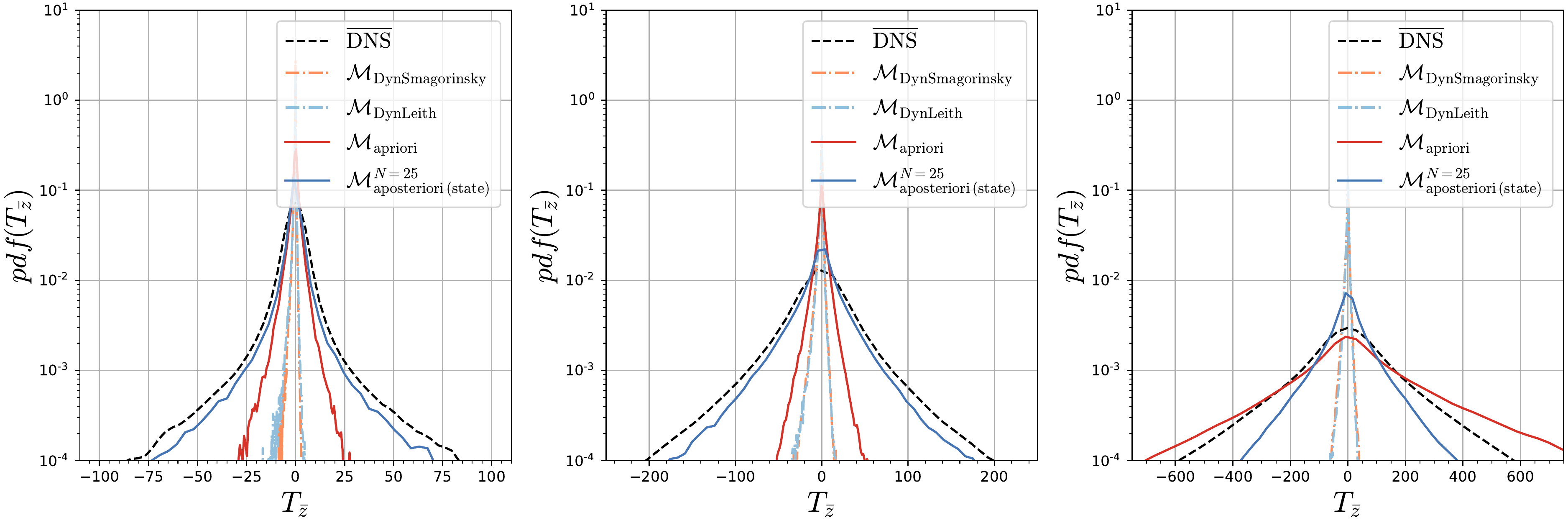}
    \caption{Probability distribution of the modeled transfer term in physical space for the three case-studies: decay (left), forced (middle) and beta-plane (right) at 
    the end of the trajectories.}
    \label{fig:pdf_transfers}
\end{figure}

\begin{figure}
    \centering
    \noindent\includegraphics[width=\textwidth]{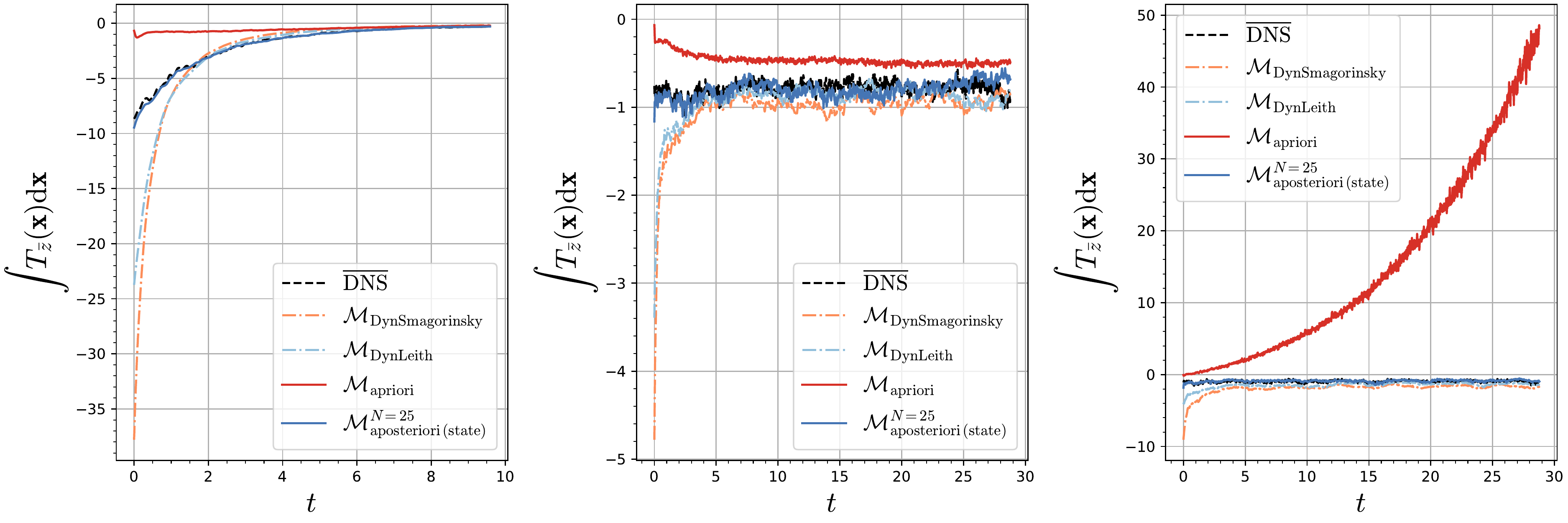}
    \caption{Temporal evolution of the spatially-averaged modeled transfer term in physical space for the three case-studies: decay (left), forced (middle) and beta-plane (right).}
    \label{fig:int_transfers}
\end{figure}

We analyze the ability of the benchmarked models to reproduce modeled transfers in physical space. 
Fig. \ref{fig:pdf_transfers} shows the PDF of $T_{\bar{z}}$ for the three different configurations at the end of their respective trajectories.
First, while we can not explicitly identify forward and backward transfers, we can see that the resolved transfer term produced by the physical models is not symmetric, i.e. it produces a larger amount of negative values. This suggests that theses models are not able to produce backscatter by construction.
The \textit{a posteriori} model always performs better than the \textit{a priori} one, with predictions that better reproduce the DNS on the tails of the distribution. 
However, even if the \textit{a priori} model has been demonstrated to be unstable on the beta-plane configuration, the resolved transfer term remains close to the DNS, but overpredicts positive tails, which, again, could indicate an incorrect representation of energy backscatter.
We can further analyze the instability related to $T_{\bar{z}}$ by looking at the evolution of its spatial average, as shown in Fig. \ref{fig:int_transfers}. 
The resolved transfer term of the \textit{a priori} model rapidly diverges from the DNS in the beta-plane configuration, which identifies the instability. 
As expected, the physical models produce large negative values at the beginning of the simulations, which can be explained by their highly dissipative nature.
The \textit{a posteriori} model, however, is the most accurate and remains close to the DNS throughout the trajectories.

\section{Discussion}
This study investigated different learning strategies to train subgrid-scale (SGS) parametrizations for two-dimensional quasi-geostrophic turbulent flows. 
While the state-of-the-art has mostly explored \textit{a priori} learning schemes, 
our numerical experiments stress the significant improvement brought by the \textit{a posteriori} learning strategy to better reproduce small-scale dynamics on large temporal horizons with great accuracy. 
For all the flow configurations and coarsening schemes considered in this study, 
SGS parametrizations trained according to an \textit{a posteriori} training loss clearly outperform both physics-based and machine learning baselines.

The \textit{a posteriori} learning strategy introduced in this paper opens the possibility to design stable subgrid parametrizations with more flexibility than state-of-the-art ML-based approaches. Indeed, we have here explored a relatively simple \textit{a posteriori} training loss given by a vorticity-based MSE, the \textit{a posteriori} learning scheme offers a much greater flexibility for the exploitation and combination of different \textit{a posteriori} metrics during the learning phase. Losses defined from classic performance metrics such as energy transfers and distributions seem particularly appealing. One may also explore application-specific metrics including among others boundary layers flows. As the \textit{a posteriori} learning strategy results in an improved stability of the trained SGS parametrizations, it may also offer means to explore more complex neural architectures for SGS terms. Here, we considered a relatively simple ConvNet, but more complex and state-of-the-art neural architectures including for instance ResNet, UNet and transformer networks could be worth exploring. Another interesting avenue is the joint training of \textit{a posteriori} models in the context of data assimilation such as described in \citeA{bonavita2020machine} and \citeA{farchi2021using}.

An interesting connection can indeed be made between \textit{a posteriori} learning and variational data assimilation techniques. Our \textit{a posteriori} learning algorithm formulates a variational problem which is formally equivalent to the strong constraint 4D-Var scheme \cite{blayo2015advanced,carrassi2018data}. But, in our case, the {\em control vector} is composed of the parameters of the neural network and observations are assumed to be perfect. The analogy between \textit{a posteriori} learning and 4D-Var therefore brings the question of whether parametrizations, or more generally corrections to existing models, could be learned directly from sparse and noisy observations \cite{schneider2017earth}. In this sense, \textit{a posteriori} learning is related to the bias correction methods that have been proposed in data assimilation \cite{dee2005bias}, and especially the schemes proposed to infer state-dependent corrections to existing models \cite{griffith2000adjoint,d2000reducing}. Interestingly, this field has received renewed attention over recent years with several authors proposing to approach bias correction with ML \cite{bonavita2020machine, farchi2021using}. In this context, we stress that our approach is very similar to the scheme introduced by \citeA{farchi2021using}, with the noticeable difference that we here learn a correction through the 4D-Var scheme itself, and not from the increment of the assimilation scheme.

By construction, parametrizations learned with the \textit{a posteriori} strategy should improve the short-term forecasting capabilities of the models. This is true in particular if the training loss is defined as the sum of forecasting errors over a given time horizon as considered here. Interestingly, we noted here that for SGS parametrizations, this short-term forecasting performance translates in a better long-term stability and representation of long-term flow patterns where the long-term horizon being several order of magnitudes greater than the time horizon in the training loss (18000 vs. 25 time steps). While recent studies have explored neural models for the short-term forecasting of realistic geophysical flows, especially for weather forecasting applications \cite{schultz2021can,weyn2021sub}, we believe our study opens new avenues for the exploitation of  learning-based components in climate-scale simulations, which remain an open challenge \cite{rasp2018deep}. In this respect, to account for the chaotic nature of turbulent flows, \textit{a posteriori} training losses could also benefit from statistical metrics as opposed to synoptic ones as the MSE used in this work.

But a strong requirement of the proposed framework lies in the differentiability of the considered dynamical model, which may question its practical applicability. 
Indeed, most large-scale forward solvers in earth system models (ESM) rely on high-performance languages that do not embed automatic differentiation (AD) capabilities. 
While it is generally recognised that adjoint models are very useful additional tools for these solvers \cite{barkmeijeradjoint2009,wunsch2013dynamically}, 
adjoint operators are readily available only for a small fraction of them \cite{heimbach2005efficient,vidard2015nemotam}. 
We stress that the emergence of a new generation of models written in differentiable programming languages such as JAX and JuliaDiff 
\cite{hafner2021fast,ramadhan2020oceananigans,sridhar2021large,huang2021jlbox} naturally supports our contribution. Besides, 
deep differentiable emulators \cite{nonnenmacher2021deep,hatfield2021building,kasim2021building} that learn a differentiable approximation of a 
non-differentiable forward solver or of its adjoint may also open new avenues for the development of SGS parametrizations for state-of-the-art ESMs with \textit{a posteriori} learning strategies.

Still, while one could benefit from the ongoing effort for differentiable ESMs, the development and the practical implementation of ML-based parametrizations in non-differentiable ESMs \cite{marshall1997finite,madec2017nemo} is also an important question. In this context, physics-based neural networks using \textit{a priori} training schemes arise as relevant options, given the promising results already reported for the oceanic \cite{zanna2021deep} and atmospheric \cite{gentine2021deep} parts of climate models. We also believe that the proposed \textit{a posteriori} strategy combined with physics-based constraints will also be worth exploring in future work.

\section*{Data Availability Statement}
The results can be reproduced from the data, models and associated learning algorithms provided along with the pseudo-spectral quasi-geostrophic code, available in \url{https://doi.org/10.5281/zenodo.6799035}.

\acknowledgments
The authors would like to thank Laure Zanna, Olivier Pannekoucke, Corentin Lapeyre and Emmanuel Cosme for helpful discussions. 
This research was supported by the CNRS through the 80 PRIME project and the ANR through the Melody, OceaniX, and HRMES ANR-17-MPGA-0010 projects. 
Additional support was also provided by Schmidt Futures, a philanthropic initiative founded by Eric and Wendy Schmidt, as part of its Virtual Earth System Research Institute (VESRI). 
Computations were performed using GPU resources from GENCI-IDRIS.
The authors also wish to thank the reviewers for their insightful suggestions that helped to improve this manuscript.


%
%

\bibliography{references}

%
%
%
%
%

\end{document}


%
%


\title{Supporting Information for "Insert Title"}
%
%

%
%



\authors{=Authors=}


\affiliation{=number=}{=Affiliation Address=}

%
%

%

\begin{article}

%
%

\noindent\textbf{Contents of this file}
\begin{enumerate}
\item Text S1 to Sx
\item Figures S1 to Sx
\item Tables S1 to Sx
\end{enumerate}
\noindent\textbf{Additional Supporting Information (Files uploaded separately)}
\begin{enumerate}
\item Captions for Datasets S1 to Sx
\item Captions for large Tables S1 to Sx (if larger than 1 page, upload as separate excel file)
\item Captions for Movies S1 to Sx
\item Captions for Audio S1 to Sx
\end{enumerate}

\noindent\textbf{Introduction}


\noindent\textbf{Text S1.}
%


\noindent\textbf{Data Set S1.} 


\noindent\textbf{Movie S1.} 


\noindent\textbf{Audio S1.} 
\end{article}
\clearpage